\documentclass[useAMS,usenatbib]{mn2e}
\usepackage{graphicx}
\usepackage{amssymb}
\usepackage{lscape}
\usepackage{ulem}
\usepackage{txfonts}

\def\kms{km~s$^{-1}$}

\def\si2{Si\,{\sc ii}}
\def\mg2{Mg\,{\sc ii}}
\def\fe2{Fe\,{\sc ii}}
\def\al2{Al\,{\sc ii}}
\def\zn2{Zn\,{\sc ii}}
\def\c2s{C\,{\sc ii}$^{\star}$}
\def\hkpc{$h_{70}^{-1}$ kpc}

\def\Ha{H$\alpha$}
\def\Hb{H$\beta$}
\def\OIII{O,{\sc iii}}
\def\OII{O,{\sc ii}}
\def\NII{N,{\sc ii}}
\def\SII{S,{\sc ii}}

\title[The cluster mass-metallicity relation]{The mass-metallicity relation 
in galaxy clusters: the relative importance of cluster membership versus local 
environment.}

\author[Ellison et al.] {Sara L. Ellison$^1$, Luc Simard$^2$, Nicolas B.
Cowan$^3$, Ivan K. Baldry$^4$, David R. Patton$^5$, 
\newauthor Alan W. McConnachie$^{1,2}$\\
$^1$ Department of Physics and Astronomy, University of Victoria, Victoria, British Columbia, V8P 1A1, Canada.\\
$^2$ National Research Council of Canada,
Herzberg Institute of Astrophysics, 5071 West
Saanich Road, Victoria, British Columbia, V9E 2E7, Canada\\
$^3$ Astronomy Dept., University of Washington, Box 351580, Seattle,
WA 98195, USA. \\
$^4$ Astrophysics Research Institute, Liverpool John Moores University,
Twelve Quays House, Egerton Wharf, Birkenhead, CH41 1LD, UK.\\ 
$^5$ Department of Physics \& Astronomy, Trent University, 
1600 West Bank Drive, Peterborough, Ontario, K9J 7B8, Canada.
}

\begin{document}

\maketitle

\begin{abstract}
Using a large (14,857), homogenously selected sample of cluster
galaxies identified in the Sloan Digital Sky Survey Data Release 4
(SDSS DR4), we investigate the impact of cluster membership and local
density on the stellar mass- gas phase metallicity relation (MZR).  We
show that stellar metallicities are not suitable for this work, being
relatively insensitive to subtle changes in the MZR.  Accurate nebular
abundances can be obtained for 1318 cluster galaxies in our sample and
we show that these galaxies are drawn from clusters that are fully
repesentative of the parent sample in terms of mass, size, velocity
dispersion and richness.  By comparing the MZR of the cluster galaxies
with a sample control of galaxies matched in mass, redshift, fibre
covering fraction and rest-frame $g-r$ colour cluster galaxies are found
to have, on average, higher metallicities by up to 0.04 dex.
The magnitude of this offset does not depend strongly on galactic
half-light radius or cluster properties such as velocity dispersion or
cluster mass.  The effect of local density on the MZR is investigated,
using the presence of a near neighbour and both two- and
three-dimensional density estimators.  For all three metrics, it is
found that the cluster galaxies in locally rich environments have
higher median metallicities by up to $\sim$ 0.05 dex than those in locally
poor environments (or without a near neighbour).  Control
(non-cluster) galaxies at locally high densities exhibit similar
metal-enhancements.  Taken together, these results show that galaxies
in clusters are, on average, slightly more metal-rich than the field,
but that this effect is driven by local overdensity and not simply
cluster membership.  
\end{abstract}

\section{Introduction}

Is it nature or nurture that governs the evolution of a galaxy?  
Understanding the relative importance of the intrinsic physical
properties of a galaxy (such as its mass, gas fraction and
morphological characteristics) versus environmental effects
has been an on-going endeavour in extra-galactic astronomy.  The 
presence of apparently fundamental scaling relations,
such as  the Tully-Fisher relation, the correlation of black hole and 
spheroid mass and the fundamental-plane (e.g. Tully \& Fisher 1977;
Ferrarese \& Merritt 2000; Faber \& Jackson 1976) indicates that
internal processes that respond to intrinsic galaxy properties
may play the principal role in galaxy evolution.  However, galaxies
clearly also respond to environmental factors.  Galaxy interactions
and mergers can cause gas inflows, morphological transformations,
trigger star formation and
ultimately lead to activity in the galactic nucleus (Barton et al.
2000; Lambas et al. 2003; Nikolic et al. 2004; Alonso et al. 2007;
Ellison et al. 2008a; Woods \& Geller 2007).  Larger scale environment
also plays a role in modulating star formation rates, which are 
systematically lower in cluster
galaxies, particularly in their cores, possibly as a result of an 
earlier starburst, or a more gradual process of gas exhaustion 
(e.g. Balogh et al. 1997, 1998, 1999; Poggianti et al. 1999; Koopmann \& 
Kenney 2004).
However, it appears that local density, on scales $<$ 1 Mpc,
may be more important
than simple cluster membership in suppressing star formation 
(e.g. Lewis et al. 2002; Gomez et al. 2003; Kauffmann et al. 2004;
Blanton \& Berlind 2007; Welikala et al. 2008; Park, Gott \& Choi 2008). 
The local density of galaxies also modulates the average stellar mass,
nuclear activity and dust content (Kauffmann et al. 2004).
Star formation, it seems, responds to both intrinsic and
external factors.

The star formation history of a galaxy is thus a record of the processes
that have influenced its evolution.  Star formation eventually leads to 
the production of metals, and a correlation between a galaxy's stellar 
mass and its metallicity has been observed out to at least $z \sim 3$
(Tremonti et al. 2004; Savaglio et al. 2005; Erb et al. 2006; Maiolino 
et al. 2008;  Hayashi et al. 2008).  Given the contributions of both intrinsic 
(`nature') and environmental (`nurture') processes to the 
rate of star formation, the scatter in the stellar mass-metallicity relation
is surprisingly small, only 0.1 dex at z $\sim$ 0.1 (Tremonti et al 2004).  
There are several interpretations for the small observed scatter.  
For example,  it may be that the mass-metallicity
relation is dominated by local processes linked to stellar mass
and that metallicity is relatively insensitive
to environmental effects.  Alternatively, environmental processes
that trigger short-lived bursts of star-formation may cause a
significant, but transient change in a galaxy's metallicity before it
returns to an equilibrium metallicity (Finlator \& Dav\'e 2008).  

A number of studies have recently revealed some clues to the relative
importance of intrinsic and environmental effects on the
mass-metallicity relation.  First, it has been shown that the
mass-metallicity relation is insensitive to galaxy morphology,
as quantified by its concentration or bulge fraction
(Tremonti et al. 2004; Ellison et al. 2008b).  Conversely, galaxies
with small half-light radii ($r_h$) or low specific star formation
rates (star formation rate per unit mass, SSFR) at a given stellar
mass have elevated metallicities (Ellison et al. 2008b).  Half-light
radius and SSFR account for a spread of up to about 0.2 dex in the
stellar mass-metallicity relation.  Environment seems to play a more
modest role.  Mouhcine et al. (2007) and Cooper et al. (2008) both
studied star-forming galaxies in the Sloan Digital Sky Survey (SDSS)
as a function of local density.  The gas-phase metallicity for
galaxies (at a given stellar mass) in the richest environments is only
$\sim$0.05 dex higher than those in the poorest environments and
contributes to $\sim$ 15\% of the total scatter in the mass-metallicity
relation.  Building on the work of Sheth et al. (2006), who found that
higher SFRs in dense environments at $z \sim 3$ lead to enhanced
chemical enrichment, Panter et al. (2008) find that the mass-weighted
average metallicity of z $>$ 0.5 galaxies depends on the number of
cluster members over smoothing scales of a few degrees.  Both
Sheth et al. (2006) and Panter et al. (2008) find that at $z<0.5$
there is little correlation between metallicity and environment.  

A more significant environmental effect is observed at much smaller
scales, e.g., for galaxies either involved in close interactions, or
those that may have recently experienced a merger.  Kewley et
al. (2006) and Ellison et al. (2008a) both found that galaxies in
close pairs are more metal-poor by approximately 0.1 dex at a given
luminosity, compared with galaxies with no near companion.  Ellison et
al. (2008a) also found that, at a given stellar mass, the close pairs
have lower metallicities by 0.05 dex, indicating that the offset in
the luminosity-metallicity relation is due to both suppressed
metallicity \textit{and} increased luminosity.  Selecting galaxies
that showed signs of recent, intense star formation activity, possibly
due to a merger, Hoopes et al. (2008) and Rupke et al. (2008) also
found depressed metallicities by up to 0.3 dex.  The interpretation of
these results is that galaxies involved in interactions experience an
inflow of metal-poor gas to their centres which triggers a burst of
star formation which in turn enriches the interstellar medium (ISM).
Hence, galaxies observed early in this sequence have low metallicities
due to metal-poor gas inflow, but once the interaction-induced star
formation is complete, the newly synthesized metals enhance the ISM
metallicity.  However, several questions remain unanswered.  For
example, what is the timescale for metallicity modulation in
interactions/rich environments?  Do galaxies in rich environments
remain metal-enhanced, or do they return to an equilibrium metallicity
for their mass, as suggested by some models (Finlator \& Dav\'e 2008)?
On what scale can local environment alter the position of a galaxy in
mass-metallicity space?

These questions can be addressed by studying the chemical abundances
of galaxies in clusters. Although the cores of clusters represent
some of the richest and most over-dense environments in the Universe,
their outskirts can be relatively low density.  Clusters also
host galaxies in a range of evolutionary phases - starbursts,
mergers and passively evolving gas-exhausted galaxies can all be found
in galaxy clusters.  Clusters therefore present the possibility to
disentangle the various processes affecting a galaxy's metallicity.
However, there have been relatively few studies of the dependence
of galaxy metallicity in clusters, and extant results paint a
confusing picture.

Skillman et al. (1996) found that HI deficient galaxies in the centre
of the Virgo cluster exhibit metallicity enhancements relative to
galaxies at the cluster periphery and field spirals of similar
luminosities by 0.3 to 0.5 dex.  The metallicity enhancement
in these late-type Virgo cluster galaxies has been confirmed by Pilyugin
et al. (2002) and Dors \& Copetti (2006).  Skillman et al.
(1996) concluded that local dynamical
processes are more important than simple cluster membership in
determining the metallicity of a galaxy.   However, the Virgo cluster
sample consists of only 9 spiral galaxies, of which 3 are HI poor.  The
field spiral galaxy sample with which they make their comparison
(Zaritsky et al. 1994) had previously been used to demonstrate that
metallicity depends not only on luminosity, but also on Hubble type
and maximum circular velocity.  Skillmann et al. (1996) indeed
conclude that `The dispersion in properties of field galaxies and the
small size of the Virgo sample make it difficult to draw definitive
conclusions about any systematic differences between the field and
Virgo spirals'.  At higher redshifts, Mouhcine et al. (2006) studied
17 massive star-forming cluster galaxies between 0.3 $< z <$ 0.6 and
found that their metallicities were consistent with the field.

There are also conflicting results for lower mass galaxies.
Vilchez (1995) and Lee, McCall \& Richer (2003) both found that
local cluster dwarfs have metallicities (at a given luminosity) consistent
with the field.  However, in a more recent study, Lee, Bell \& Somerville
(2008) found that stellar metallicities tend to be lower at a given
baryonic mass for cluster dwarfs than their field sample.  

The main limitation of these studies has been a statistical one, 
with insufficient numbers of galaxies
available to fully dissect possible trends from the many interconnected
parameters of galaxy evolution.    In this paper, we tackle the 
question of environmental metallicity dependence
by using the statistical power of the SDSS.  Our method fundamentally
differs from that of Mouhcine et al. (2007) and Cooper et al. (2008),
who look for offsets in the mass-metallicity relation as a function
of local density.  The approach taken here is to assemble a sample
of known cluster galaxies and to compare them to a control
sample that is matched in basic physical properties.  This distinction
is important in assessing whether it is simply the local
density of galaxies that influences chemical enrichment, or
whether cluster membership is important.  For example, Poggianti
et al. (2008) find that even though the fraction of post-starburst
galaxies is nearly independent of local density, post-starbursts
preferentially reside in massive clusters.  This was interpreted
as evidence that the intra-cluster medium plays an important
role in shutting down star formation in these galaxies.

The paper is organised as follows.  Sections \ref{C4_sample_sec}
through \ref{control_sample_sec} describe the pre-analysis steps
of sample selection, metallicity diagnostic choice and the
characterization of galaxy cluster properties.  Specifically,
in Section \ref{C4_sample_sec}
the selection of the cluster galaxies from the SDSS DR4 is described.
Both stellar and nebular metallicities are available for this
sample and in Section \ref{met_sec} we describe the advantages and
disadvantages of these two diagnostics and argue that the gas-phase
abundances are best suited to this study. Section \ref{control_sample_sec}
describes the compilation of the control sample to which
the cluster galaxy mass-metallicity relation will be compared.
The main science results of this paper are presented in Sections
\ref{C4_MZR_sec} (dependence of the mass-metallicity relation on
cluster membership) and \ref{env_sec} (dependence of the mass-metallicity 
relation on local environment).

\begin{figure}
\centerline{\rotatebox{0}{\resizebox{8cm}{!}
{\includegraphics{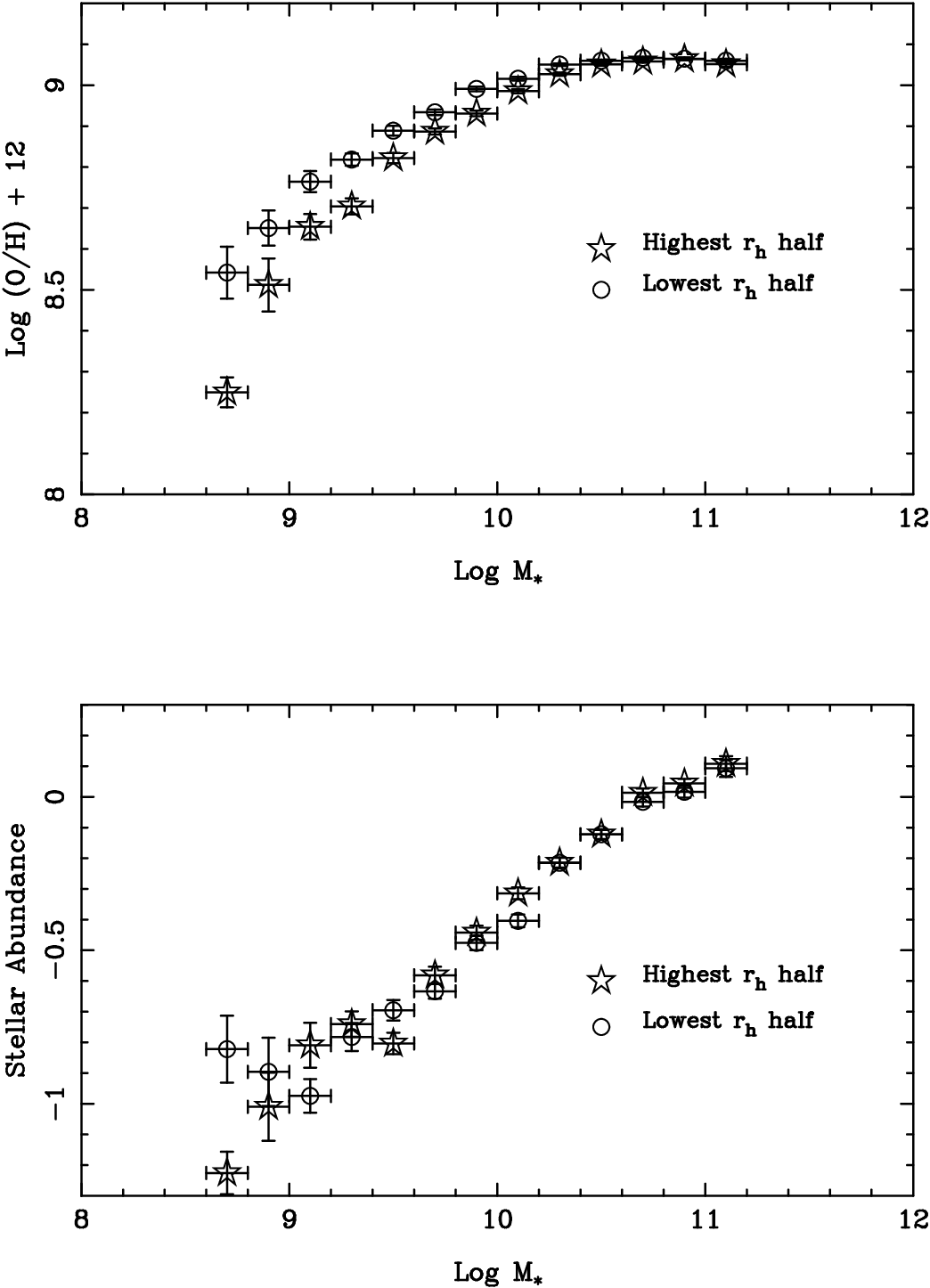}}}}
\caption{ Field galaxies show sequences in the stellar mass nebular
metallicity relation (upper panel) according to their half-light radius, 
as shown by Ellison et al. (2008b).  There is no such clean cut
sequence for the stellar mass stellar metallicity relation
of field galaxies (lower panel).}
 \label{MZ_rh_field}
\end{figure}

\section{Selection of the mass-metallicity sample}\label{C4_sample_sec}

\begin{figure*}
\centerline{\rotatebox{270}{\resizebox{12cm}{!}
{\includegraphics{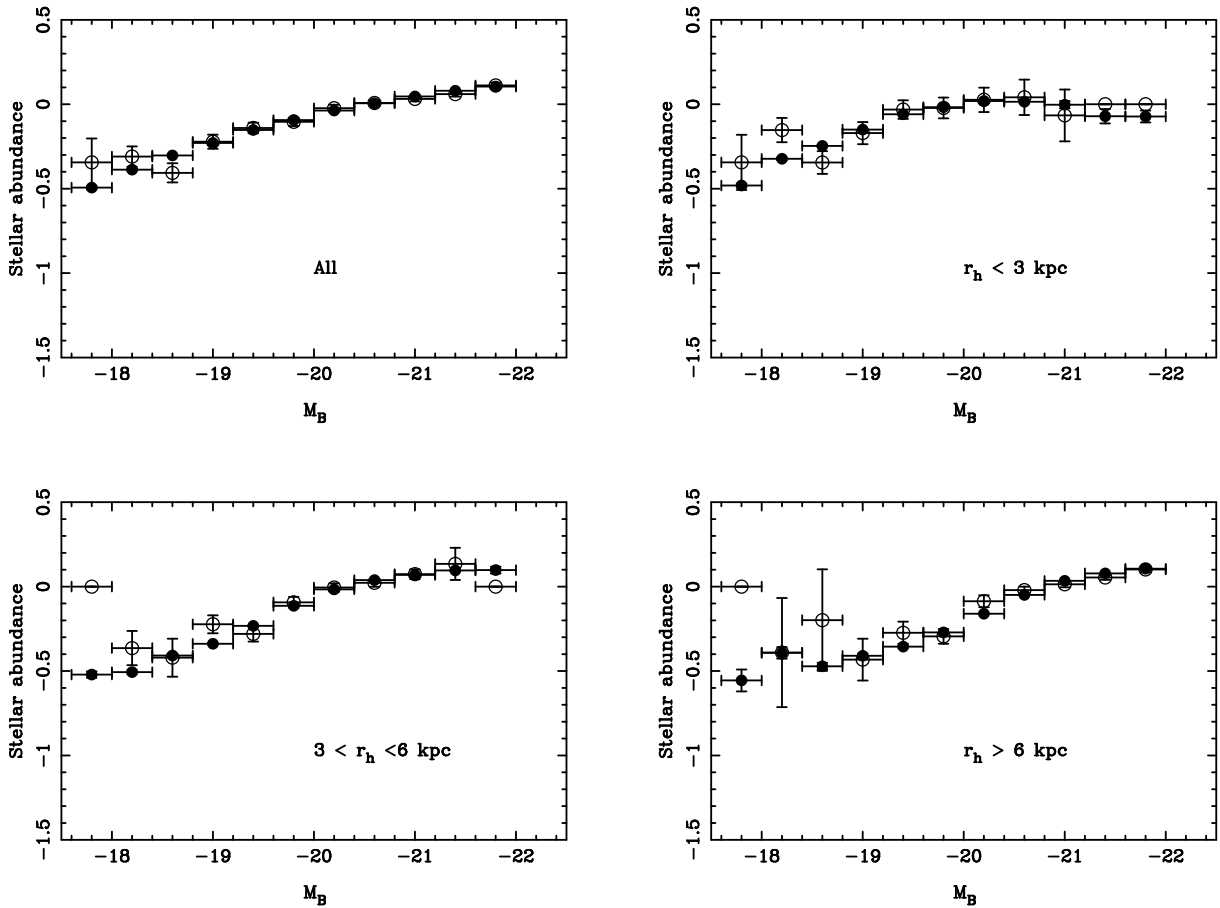}}}}
\caption{The luminosity-stellar metallicity relation for galaxies
with close companions ($\Delta v < 500$ \kms\ and $r_p < 30$ \hkpc,
open circles) compared with a control sample (solid circles).
See Ellison et al. (2008a) for full details of the pair sample.
Unlike the stellar mass-nebular metallicity relation for galaxy
pairs (see Figure 15 of Ellison et al. 2008a) there is no
clear offset between the pairs and the control.
\label{LZ_pairs} }
\end{figure*}

The cluster galaxies in our sample have been identified using the
C4 algorithm which selects cluster members based on a seven parameter
assessment of both position and colour.  The algorithm was applied to
the SDSS DR2 by Miller et al. (2005) and subsequently extended to
later data releases (e.g. von der Linden et al. 2007).  Miller et al.
(2005) estimate that the DR2 C4 cluster sample is approximately 90\% 
complete and 95\% pure.  von der Linden et al. (2007) critically
re-assessed the C4 catalogue, improving the cluster velocity
dispersions and redshifts.  Following these refinements, von der
Linden et al. (2007) constructed a sample of 625
clusters from the SDSS DR4 for which the brightest 
cluster galaxy (BCG, $z \le 0.1$) can be confidently identified.  
These 625 clusters, containing a total of 18,100 galaxies, represent
the parent sample from which we compile our list of cluster galaxies. 
In order to be considered in our final cluster sample, we further require
that:

\begin{enumerate}

\item  The extinction-corrected
Petrosian $r$-band magnitude is in the range 14.5 
$< r <$ 17.77.  

\item Redshifts are available and that the SDSS redshift quality 
control flag $z_{\rm conf} > 0.7$.

\item  Objects are unique and have a SciencePrimary flag = 1 (to remove 
spectroscopic duplicates).

\item  Galaxies are assigned to a unique cluster.  In the original
C4 sample of 18,100, 1179 galaxies have been assigned to more than one 
cluster. 

\item  Galaxy stellar masses are available in Kauffmann et al. (2003)
catalogue.

\item  The fibre covering fraction (CF) is at least 20\% in the $g$-band 
in order to avoid aperture effects.  

\item Both $g$ and $r$ band magnitudes are available.  
These are converted to the rest frame using
k-corrections from Blanton \& Roweis (2007).

\end{enumerate}

These cuts yield a sample of 14,857 galaxies in what we shall
refer to as the refined C4-DR4 catalogue.  Only a subset of these
galaxies have some reliable measurement of metallicity.  In the following
section we describe our selection of those galaxies and the technique
that is used to quantify the galaxy metallicity.

\section{Stellar vs. nebular metallicities}\label{met_sec}

\begin{figure*}
\centerline{\rotatebox{270}{\resizebox{12cm}{!}
{\includegraphics{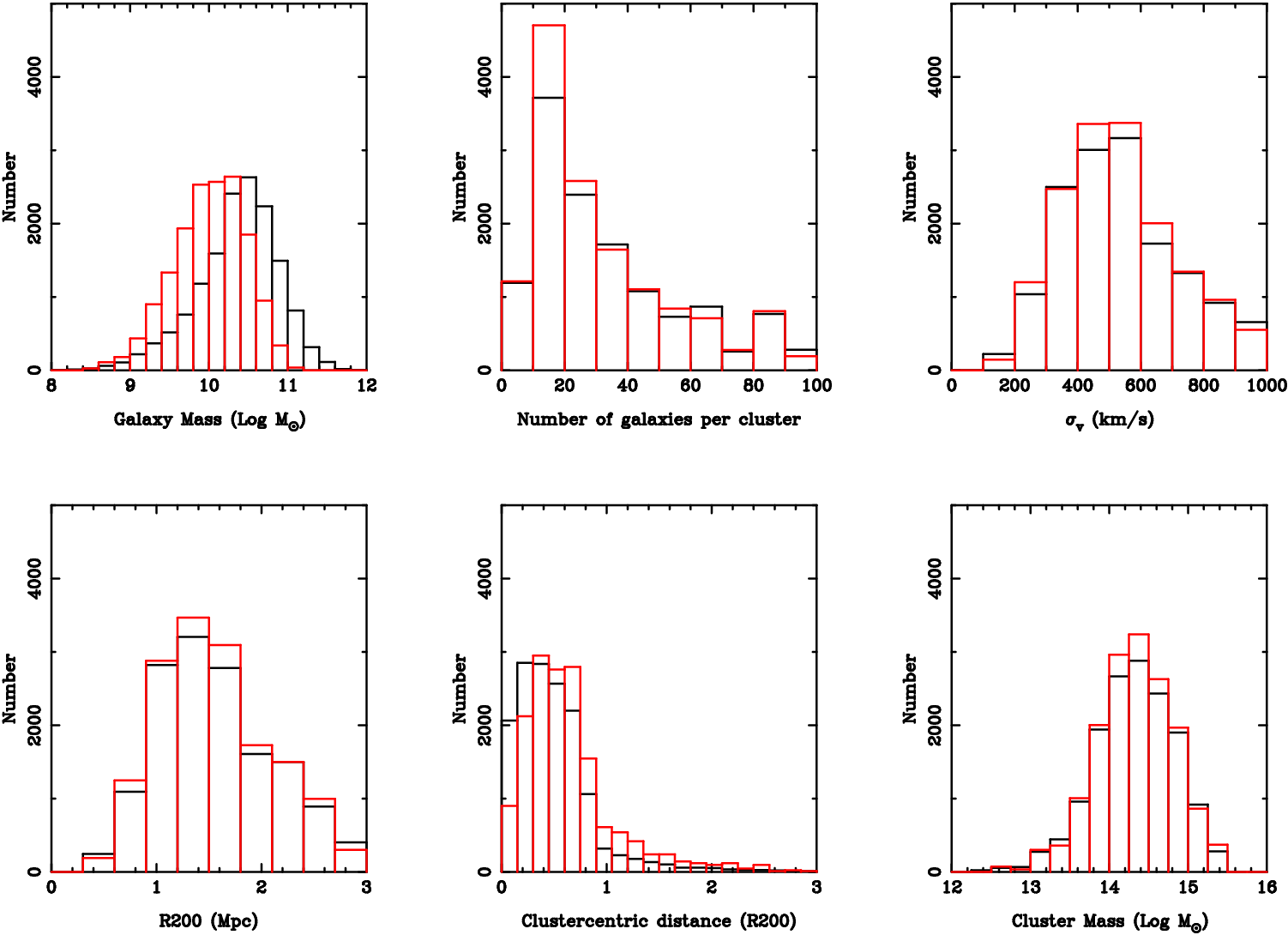}}}}
\caption{Properties of galaxies and their host clusters for the
refined C4-DR sample (black) and the subset with reliable nebular
metallicities (red), where the latter histograms are scaled by a factor
 of 11 for presentation purposes.  
The mean values of these histograms are given
in Table \ref{neb_ste_cluster_props}.  Although the clusters hosting
galaxies with nebular metallicities are indistinguishable from
the refined C4-DR4 sample in terms of $\sigma_v$, cluster mass and
R200, they contain slightly fewer galaxies on average, have a lower
mean stellar mass and tend to be located further from the cluster centre. 
\label{galaxy_metals_neb} }
\end{figure*}

\begin{figure*}
\centerline{\rotatebox{270}{\resizebox{12cm}{!}
{\includegraphics{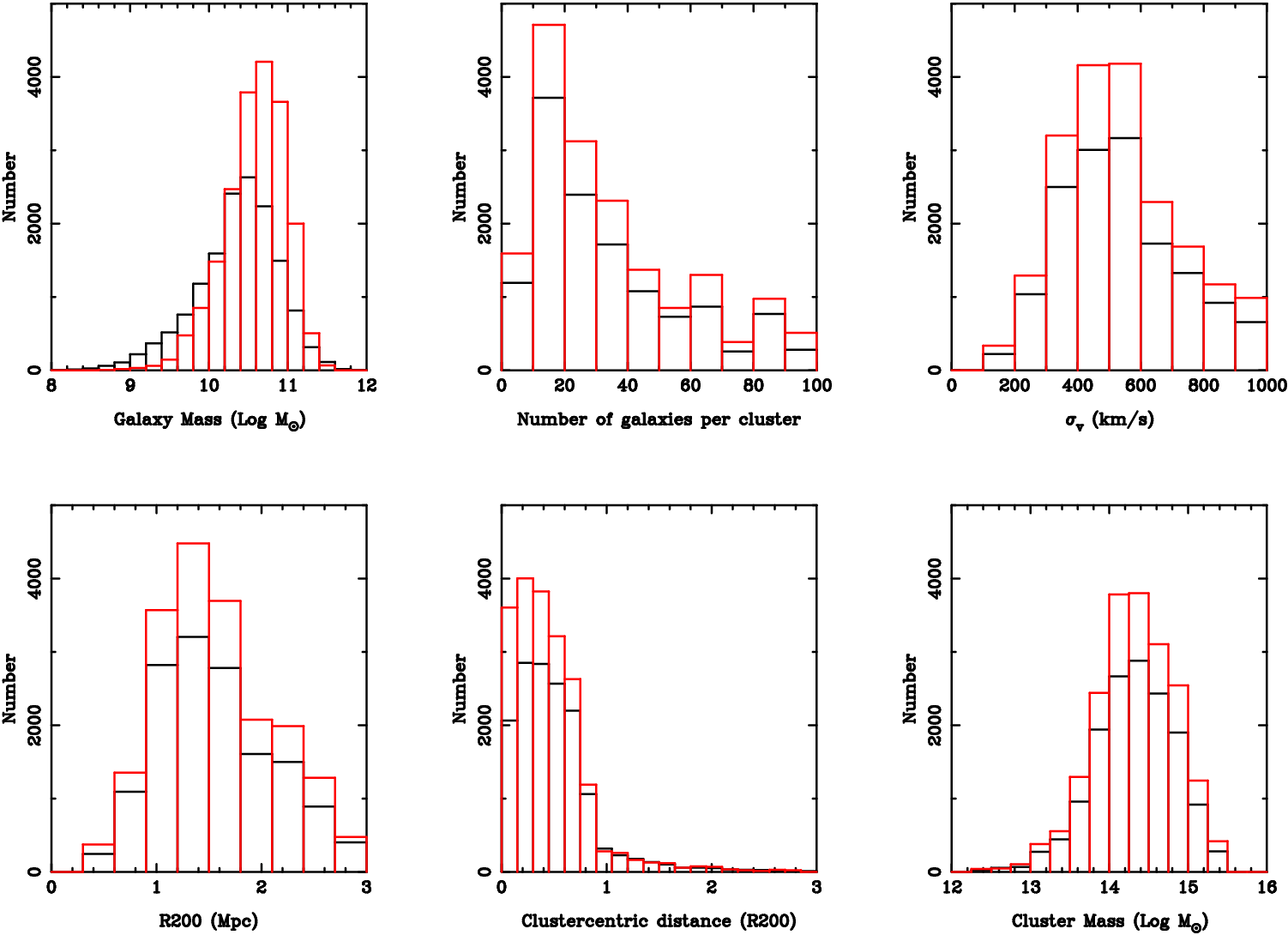}}}}
\caption{Properties of galaxies and their host clusters for the
refined C4-DR sample (black) and the subset with reliable stellar
metallicities (red), where the latter histograms are scaled by a factor
 of 4 for presentation purposes. 
The mean values of these histograms are given
in Table \ref{neb_ste_cluster_props}.  Although the clusters hosting
galaxies with stellar metallicities are indistinguishable from
the refined C4-DR4 sample in terms of $\sigma_v$, cluster mass and
R200, they contain slightly more galaxies on average, have a higher
mean stellar mass and tend to be located closer to the cluster centre. 
\label{galaxy_metals_ste} }
\end{figure*}

\begin{center}
\begin{table*}
\caption{Properties of the refined C4-DR4 galaxies and the subsets from
which reliable nebular and stellar metallicities can be determined.}
\begin{tabular}{lccc}
\hline
~~~~~~~~~~ & All refined C4-DR4 galaxies & C4-DR4 Galaxies with nebular metallicities
& C4-DR4 Galaxies with stellar metallicities \\ 
\hline
Number of galaxies & 14,857 & 1318 & 3952 \\
Mean redshift & 0.068 & 0.070 & 0.065\\
Mean \# galaxies/cluster & 49.1 & 45.4 & 50.9\\
Mean $\sigma_v$ (\kms) & 543.5 &  544.2 & 546.4 \\
Mean R200 (Mpc) & 1.61 & 1.61 & 1.61\\
Mean RR200 (R200) & 0.50 & 0.64 & 0.49\\
Mean cluster mass (log M$_{\odot}$) & 14.3 & 14.3 & 14.3 \\ 
Mean galaxy M$_{\star}$ (log M$_{\odot}$) & 10.4 & 10.0 & 10.6\\  
\hline 
\end{tabular}\label{neb_ste_cluster_props}
\end{table*}
\end{center}

Almost all extant work on the global stellar mass-metallicity relation of 
galaxies uses nebular metallicities derived from strong emission
lines (e.g. Tremonti et al. 2004; 
Ellison et al., 2008a,b;  Michel-Dansac et al. 2008;   Peeples et al. 2008),
including previous work on the mass-metallicity relation as a 
function of environment
(Mouhcine et al. 2007;  Cooper et al. 2008).  Although there is
a large scatter for individual galaxies, stellar metallicities
statistically trace nebular metallicities, leading to a similar
relation between stellar mass and stellar metallicity (Gallazzi et al. 
2005).  A number of authors have noted the tendency for
stellar metallicities to be offset to lower values than nebular
metallicities  (e.g. Cid-Fernandes et al. 2005; Gallazzi et al. 2005; Asari 
et al. 2007; Panter et al 2008; Halliday et al. 2008).  This is usually
attributed to the fact that stellar metallicities trace the
aggregate past of metal enrichment, whereas nebular metallicities
contain a significant contribution of metals from the most recent
star formation episodes.  A further factor may be the calibration
of nebular metallicities.  Kewley \& Ellison (2008) have shown that,
depending on the choice of strong line nebular abundance diagnostic,
the metallicity at a given stellar mass can vary by up to
0.7 dex (a factor of 5).
In this section we consider whether stellar or nebular metallicities
are best suited to our investigation of the cluster stellar mass-metallicity
relation.  Our choice is driven by requiring both sensitivity to
environmental effects and statistical leverage (sample size).

In the first instance, it may appear that stellar metallicities
are the better choice for investigating galaxies in clusters, due
to the high fraction of passively evolving galaxies therein.
Gallazzi et al. (2005) showed that despite a large scatter between
the nebular and stellar metallicities of individual galaxies, an
overall relation between stellar mass and stellar metallicity does
exist, albeit with larger scatter than the stellar mass-nebular metallicity
relation of Tremonti et al. (2004).  Since the offset in the mass-metallicity
relation due to environment may be comparable to the scatter (e.g.
Ellison et al 2008a,b;  Cooper et al. 2008)
we begin by assessing whether known trends in the stellar mass-metallicity
relation can be recovered when
stellar metallicity is used instead of nebular metallicity.  In the
following tests we use the nebular metallicities of SDSS DR4
galaxies determined from the Kewley \& Dopita (2002) `recommended'
calibration and
stellar metallicity from Gallazzi et al. (2005).  In order to be
considered in our metallicity analysis, we require the following
additional critera:

\begin{enumerate}

\item A median spectral S/N of 20 per pixel for stellar abundance
  determinations (Gallazzi et al. 2005).  For the calculation of
  nebular abundances, we require that the S/N in the emission lines of
  \OII~$\lambda \lambda3726,9$, \Hb, \OIII~$\lambda 5007$, \Ha,
  \NII~$\lambda 6584$, and \SII~$\lambda \lambda 6717,31$ is greater
  than 5.

\item Classification as an HII region dominated galaxy for nebular abundance
determination.  We use the Kewley et al. (2001) line ratio classification 
scheme to determine which galaxies have an AGN component.

\end{enumerate}

Ellison et al. (2008b) showed that the stellar mass-nebular metallicity
relation is separated into clear trends of half-light radius and
specific star formation rate.  
We note in passing that this same sequence
in half-light radius is present in our C4 cluster sample as in
the field. These sequences were interpreted by Ellison et al. (2008b)
as a result of differing star formation efficiencies.  In Figure
\ref{MZ_rh_field} the sample of field galaxies from Ellison
et al. (2008b) is used 
to determine the stellar mass-stellar metallicity relation
as a function of $r$-band half light radius.  In order to compare
the result with that of Ellison et al. (2008b) we have selected
those galaxies for which both stellar and nebular abundances can
be reliably determined (see the criteria above).   
We plot the mass-metallicity relation for this
sample of 6100 galaxies in Figure \ref{MZ_rh_field}, divided into
two bins of $r_h$.  Following Ellison et al. (2008b), we calculate
the median value of $r_h$ for each mass bin.  Figure \ref{MZ_rh_field}
confirms the dependence of the stellar mass - nebular metallicity
relation on $r_h$, but also shows that no such dependence is
present in the stellar mass - stellar metallicity relation.

The stellar mass-nebular metallicity relation in close pairs of
galaxies has recently been studied by Ellison et al. (2008a)
and Michel-Dansac et al (2008).  Both papers find that close
pairs of galaxies have a systematically lower metallicity at a given
stellar mass than a control sample, by about 0.05 dex (with the exception
of the less massive galaxy in disparate mass interactions which may have
an enhanced metallicity).  Ellison
et al. (2008a) also showed that this offset is dependent on
half-light radius, with the smallest paired galaxies exhibiting the
largest offset to low metallicities.  A similar
effect is seen in the luminosity-nebular metallicity relation,
although with a slightly larger metallicity offset ($\sim 0.1$ dex; 
Kewley et al. 2006;
Ellison et al. 2008a) indicating that changes in both metallicity
and luminosity are at work.  In Figure \ref{LZ_pairs} we investigate
whether the offset in the luminosity - metallicity relation of
paired galaxies can be recovered from stellar metallicities.  
The sample of pairs includes galaxies with close companions
within a factor of 10 in stellar mass, with velocity separations $\Delta 
v < 500$ \kms\ and projected separation $r_p < 30$ \hkpc\
(see Ellison et al. 2008a for further details on the selection
of the pairs sample).  The luminosity metallicity relation is shown
here since the metallicity
offset is larger than for the mass-metallicity relation.  However,
Figure \ref{LZ_pairs} shows that there is no offset in the luminosity -
stellar metallicity relation for galaxy pairs, relative to a control
sample.  

\subsection{C4 galaxy properties in the mass-metallicity sample}\label{MZ_sample_sec}

\begin{figure*}
\centerline{\rotatebox{270}{\resizebox{12cm}{!}
{\includegraphics{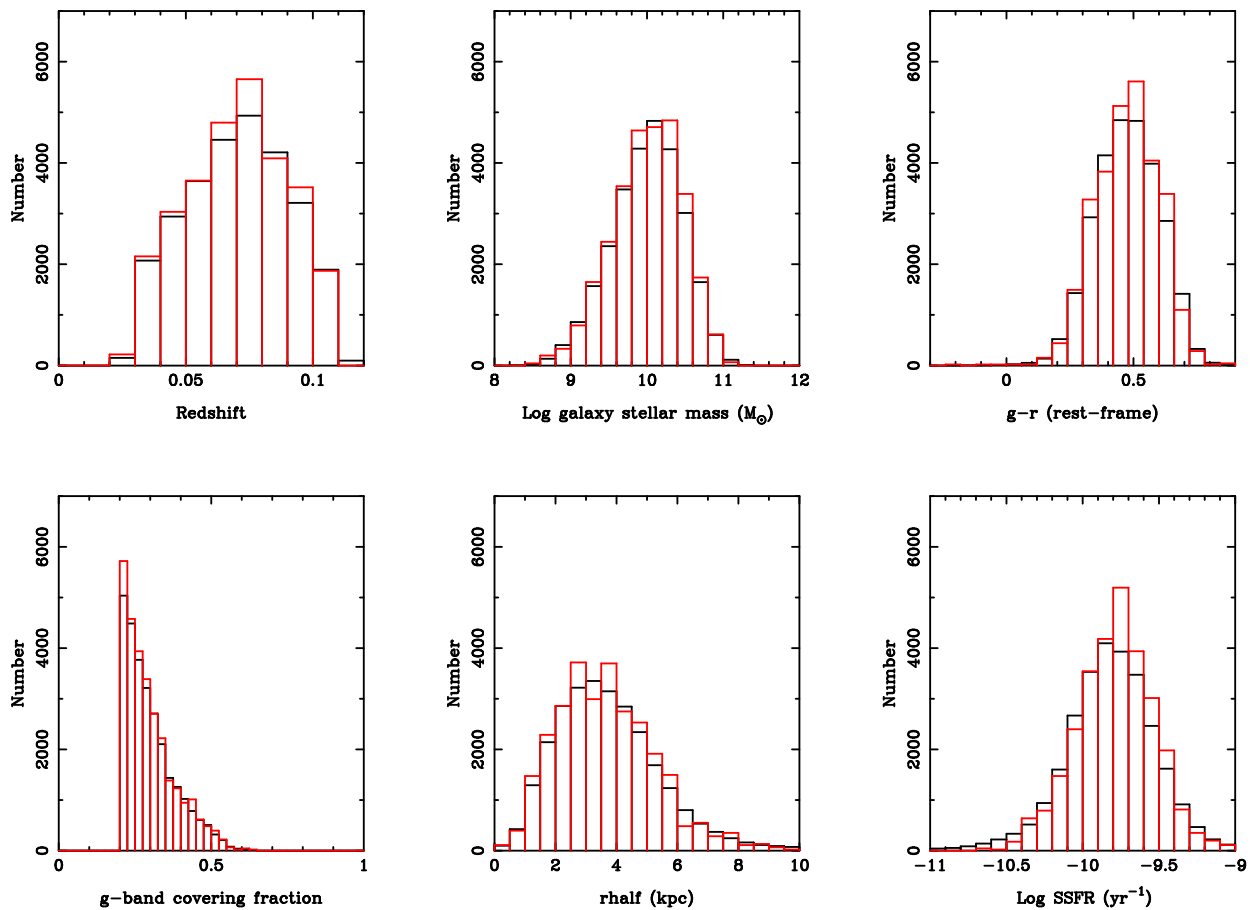}}}}
\caption{The control sample is constructed by iteratively matching
the C4-MZ cluster galaxies to SDSS DR4 non-cluster galaxies in
redshift, stellar mass, covering fraction and rest frame $g-r$ colour.
There are 21 control galaxies for each of the 1318 C4-MZ cluster galaxies.
This figure shows the distribution of the four matched parameters
for the C4-MZ cluster galaxies (red) and their control (black),
galaxies scaled for presentation purposes.\label{compare_control} }
\end{figure*}

\begin{center}
\begin{table*}
\caption{Properties of C4-MZ galaxies and their control galaxies}
\begin{tabular}{lccc}
\hline
~~~~~~~~~~ & C4-MZ cluster galaxies & Matched control galaxies & KS prob \\ 
\hline
Number of galaxies & 1318 & 27678 & N/A\\
Mean redshift & 0.070 & 0.070 & 75\% \\
Mean stellar mass (M$_{\odot}$) & 10.0 & 10.0 & 52\%\\
Mean rest-frame $g-r$ & 0.47 & 0.47 & 40\% \\
Mean $g$-band CF & 30\% & 30\% & 70\%\\
Mean $g$-band r$_h$ (\hkpc) & 4.15 & 4.02 & 96\% \\
Mean log SSFR (yr$^{-1}$) & $-9.70$ & $-9.71$ & 0.001\% \\
\hline 
\end{tabular}\label{tab_compare_control}
\end{table*}
\end{center}

\begin{figure*}
\centerline{\rotatebox{270}{\resizebox{12cm}{!}
{\includegraphics{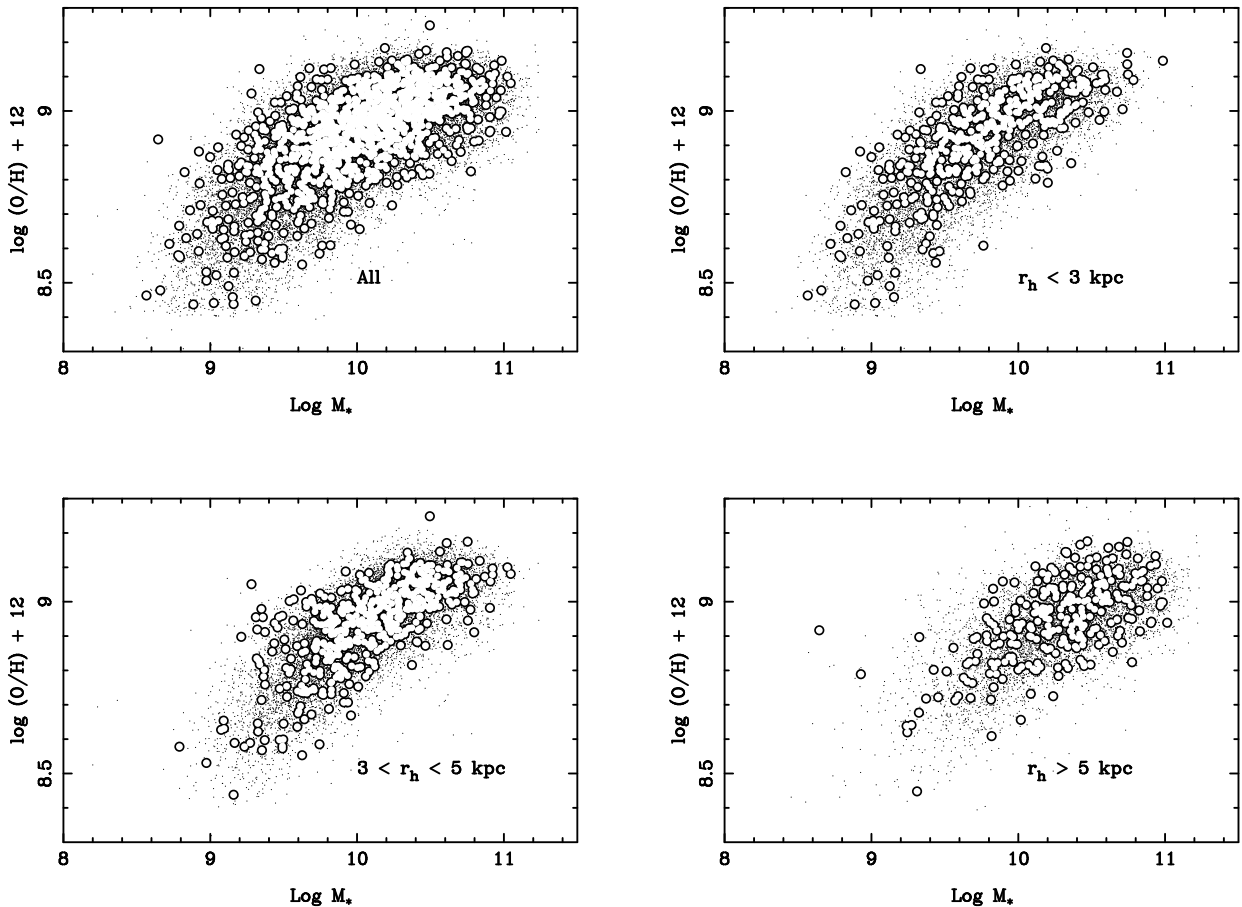}}}}
\caption{The stellar mass-metallicity relation for C4-MZ cluster
galaxies (open circles) and their control galaxies (dots)
for different cuts in half light radius.  There are 21 control galaxies
for each C4-MZ cluster galaxy.
\label{plot_MZ_unbin_rh} }
\end{figure*}

\begin{figure*}
\centerline{\rotatebox{270}{\resizebox{12cm}{!}
{\includegraphics{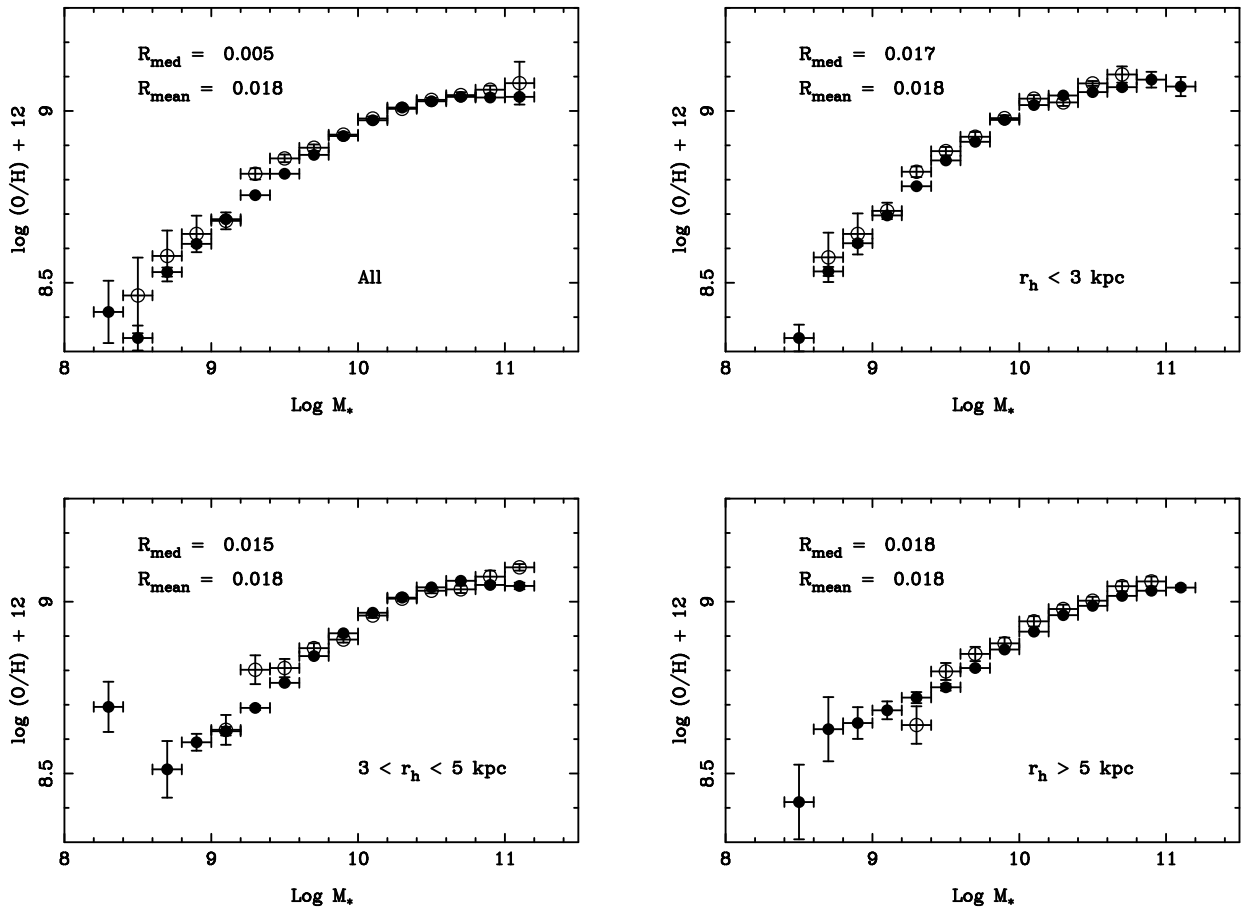}}}}
\caption{The stellar mass-metallicity relation for C4-MZ cluster
galaxies (open circles) and their control galaxies (solid circles)
for different cuts in half light radius, binned by mass.
\label{plot_MZ_bin_rh} }
\end{figure*}

The results of the previous subsection indicate that stellar metallicity 
is not very sensitive to subtle dependences on galaxy characteristic
or environment.  The nebular metallicity is therefore adopted for
the remainder of this paper.  What impact will this have on
the number of galaxies that can be studied, and what biases may be 
introduced?  For example, Cooper et al. (2008) showed that galaxies
in the DR4 with nebular abundances are biased towards less rich
environments than the full DR4 sample.  

In Figures \ref{galaxy_metals_neb}
and \ref{galaxy_metals_ste} we show the properties of clusters and
their galaxy members for those galaxies in the refined C4-DR4
catalogue which have reliable (in terms of HII classification, S/N and
CF requirements) nebular and stellar metallicities, respectively.
There are 1318 galaxies in the refined C4-DR4 with nebular metallicites,
compared with 3952 with stellar metallicities.
In both figures, the distribution of properties is compared with
the $\sim$ 15,000 galaxies in the refined C4-DR4 catalogue.  Table
\ref{neb_ste_cluster_props} lists the mean values for the galaxy and
cluster properties of these various samples.   Table 
\ref{neb_ste_cluster_props} and Figures \ref{galaxy_metals_neb}
and \ref{galaxy_metals_ste} show that galaxies with available nebular
or stellar metallicities are located in clusters that
span the full range of cluster properties,
as measured by velocity dispersion ($\sigma_{v}$), cluster size (as
measured by the virial radius, R200\footnote{von der Linden et al. (2007) 
calculate $\sigma_v$ within R200 using an iterative process that also 
yields the cluster redshift.}) and cluster mass.  
Moreover, Kolmogorov-Smirnov (KS) tests show that the
metallicity subsamples are statistically indistinguishable from the
refined C4-DR4 cluster galaxy sample in these three properties.
However, cluster galaxies with nebular abundances have a significant
tendency to belong to clusters with fewer members; the inverse is
true for galaxies with stellar metalllicities.
Figures \ref{galaxy_metals_neb} and \ref{galaxy_metals_ste} also
show that galaxies with available nebular or stellar metallicities
inhabit different regions in their host cluster and have different
stellar mass distributions compared with the full refined C4-DR4 sample.
Specifically, galaxies with nebular abundances tend to have lower
stellar masses and are rarely located in the centre of the cluster
(as measured by RR200, the distance from the cluster centre in units
of R200.
Conversely, galaxies with stellar metallicities have a higher
mean stellar mass than all the mean of the full refined C4-DR4 sample and are
preferentially located near the cluster centre.

From this exercise we conclude three things:

\begin{itemize}
\item There is a sufficient number of galaxies with either nebular
or stellar metallicities to make a statistical comparison with a
control sample.

\item Galaxies with either nebular or stellar metallicities can be
used to probe clusters with the full range of properties, such
as $\sigma_v$, cluster mass and R200.

\item Selecting galaxies with either reliable stellar or reliable
nebular metallicities results in a sample that is biased relative to
the C4 sample as a whole.
\end{itemize}

However, since we have also shown that stellar metallicity is
relatively insensitive to subtle changes in metallicity, we
adopt the gas-phase oxygen abundance as our metric of metallicity.
The sample of 1318 galaxies in the refined C4-DR4 sample with reliable
nebular abundances is henceforth referred to as the C4 mass-metallicity 
(C4-MZ) sample, and represents the sample that we will use to investigate 
the effect of environment on the stellar mass-metallicity relation.
For brevity, the stellar mass - nebular metallicity relation
is abbreviated to simply `mass-metallicity relation', or MZR,
for the rest of this paper.
The results of this subsection indicate that although this sample
can probe a representative cross-section of C4 clusters, there is
a slight bias towards lower mass clusters (see Table 
\ref{neb_ste_cluster_props}) and that this sample underrepresents
cluster core environments.

\section{The control sample}\label{control_sample_sec}

In order to investigate the MZR of
the C4-MZ sample,  a control sample must be constructed for comparison.
Galaxies
in the SDSS DR4 may be considered as control galaxies if they pass the same
criteria as the C4-MZ sample (see Sections \ref{C4_sample_sec} and
\ref{MZ_sample_sec}), but are not members of the C4 catalogue.

Galactic stellar mass, luminosity and colour are themselves dependent on
environment (e.g. Kauffmann et al 2004; Baldry et al. 2006; Park
et al. 2007).  Cooper et al. (2007) also show that metallicity is
strongly correlated with colour and luminosity and point out
any correlation between local density and metallicity may simply be 
a manifestation of these inter-dependences.  We therefore construct
a control sample that is matched to the cluster sample in
galactic properties in order to detect any underlying dependence
of metallicity on environment.
We construct the control sample in the following way.  A `pool'
of candidate control galaxies is constructed from the SDSS DR4
which fulfill the same basic requirements enumerated in Section
\ref{C4_sample_sec}, but are not in the cluster sample.
For each
galaxy in the C4-MZ sample we find the control galaxy that is
best matched in stellar mass, redshift, rest-frame $g-r$ colour and CF.
Control galaxies can only be selected once; having been identified
as a match, they are removed from the potential control `pool'.
Once a match has been made for every one of the 1318 C4-MZ galaxies,
a KS test is performed to confirm that the distribution of each the
four matched properties is statistically indistinguishable
between the C4-MZ sample and the control sample.
If the KS probability is $>$ 30\%, the matching process is repeated.
This process is iterated until the KS probability drops below  30\%.
In this way, 21 control galaxies are selected for each
C4-MZ galaxy (i.e. a total of 27,678) before the KS probability
becomes unacceptable.  Figure \ref{compare_control}
shows the distribution of galaxy properties for the C4-MZ cluster
galaxies and the control sample.  Table \ref{tab_compare_control}
lists the properties of the two samples and the KS probabilities that
the two samples are the same.  Given the impact discussed above
of r$_h$ and SSFR on the mass-metallicity relation, we also compare
(although we do not match for) these two properties between the
control and C4-MZ samples.  The C4-MZ galaxies have a slightly larger
mean r$_h$ and SSFR (derived from the aperture-corrected SFRs of
Brinchmann et al. 2004).  The half-light radii
have a high KS probability of being drawn from the same population,
even though they were not matched explicitly.  On the other hand,
the SSFRs of the two samples
are statistically inconsistent at a high level.  Figure \ref{compare_control}
shows that this statistical inconsistency is due to a tendency
for marginally higher SSFRs in our cluster galaxy sample.  However,
the effect is small, as shown in Table \ref{tab_compare_control} the
median SSFR of the C4-MZ sample differs from the control by only
0.01 dex.  

\section{The stellar mass-metallicity relation for C4 cluster 
galaxies}\label{C4_MZR_sec}

Having selected those C4 cluster galaxies with reliable metallicities
(the 1318 C4-MZ galaxies) and compiled a matched sample of 
27,678 control galaxies, the MZR in the cluster environment
can now be investigated.
In Figures \ref{plot_MZ_unbin_rh} and \ref{plot_MZ_bin_rh} the
unbinned and binned MZR for the 1318 C4-MZ galaxies and their 
control galaxies are shown.
The offset between the C4-MZ and control samples is quantified
with the parameters R$_{\rm med}$ and  R$_{\rm mean}$: the median
and average difference between cluster and control bins 
over the mass range 9 $<$ log M$_{\star} <$ 11 M$ _{\odot}$.
Positive values of R$_{\rm med}$ and  R$_{\rm mean}$ indicate that
cluster galaxies are more metal-rich than the control.
Since half-light radius seems to play an important role in both
pairs (Ellison et al. 2008a) and field (Ellison et al. 2008b)
galaxies, we also plot the C4 mass-metallicity relation for 
3 cuts in r$_h$.  There is a marginal offset towards higher median
metallicities
by up to 0.04 dex for all r$_h$ cuts, with no strong dependence
on half-light radius.  This is similar
to the small positive offset in metallicity in the highest density
environments studied by Mouhcine et al. (2007) and Cooper et al.
(2008).  The small offset towards higher metallicities in the C4-MZ
sample relative to the control can not be explained by the difference
in SSFR between the two samples noted in Section \ref{control_sample_sec}.
As shown in Section \ref{MZ_sample_sec}, the SSFRs of C4-MZ galaxies 
have slightly higher
SSFRs than the control sample (see also Figure \ref{compare_control}).
However, according to Ellison et al. (2008b), galaxies with higher
SSFRs tend to have \textit{lower} metallicities for a given mass.

Ellison et al. (2008a) showed that the metallicity offset
in their pairs study is greater in the luminosity-metallicity
plane than in mass-metallicity.  This was interpreted 
as evidence that changes in both luminosity and metallicity are
occuring during close galactic passes.    We find that the offset 
in the luminosity-metallicity relation for the C4-MZ
galaxies is consistent with that in the MZR, indicating that
the star formation event that led to the now metal-enriched ISM
is no longer significantly enhancing the magnitude of the galaxy.

\begin{figure*}
\centerline{\rotatebox{270}{\resizebox{12cm}{!}
{\includegraphics{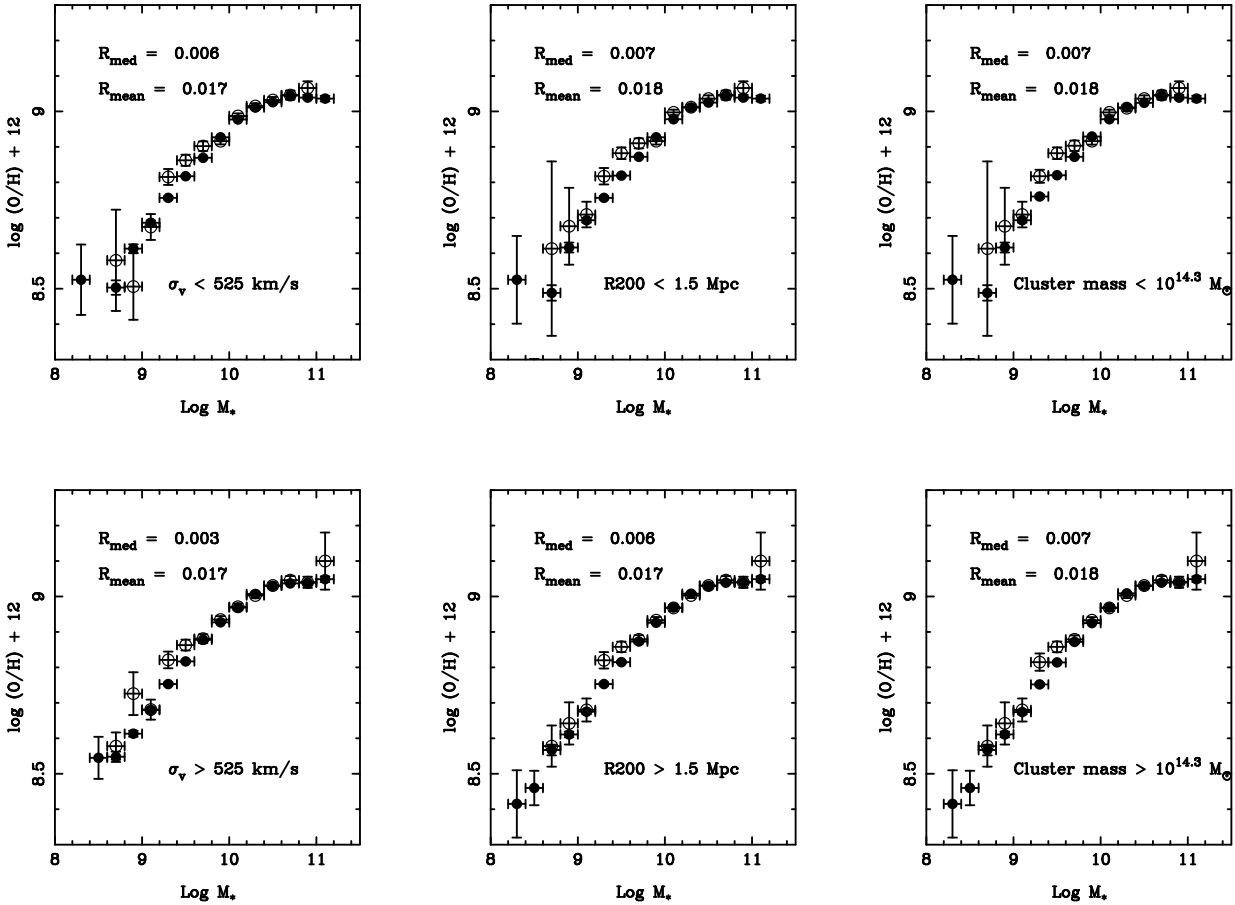}}}}
\caption{The stellar mass-metallicity relation for C4-MZ cluster
galaxies (open circles) and their control galaxies (solid circles).
The upper panels show the relationships for galaxies residing in 
low $\sigma_v$, low R200 and low mass clusters.  The lower panels
show the relationships for galaxies residing in 
high $\sigma_v$, high R200 and high mass clusters.
\label{plot_MZ_cluster_prop} }
\end{figure*}

\begin{figure}
\centerline{\rotatebox{0}{\resizebox{9cm}{!}
{\includegraphics{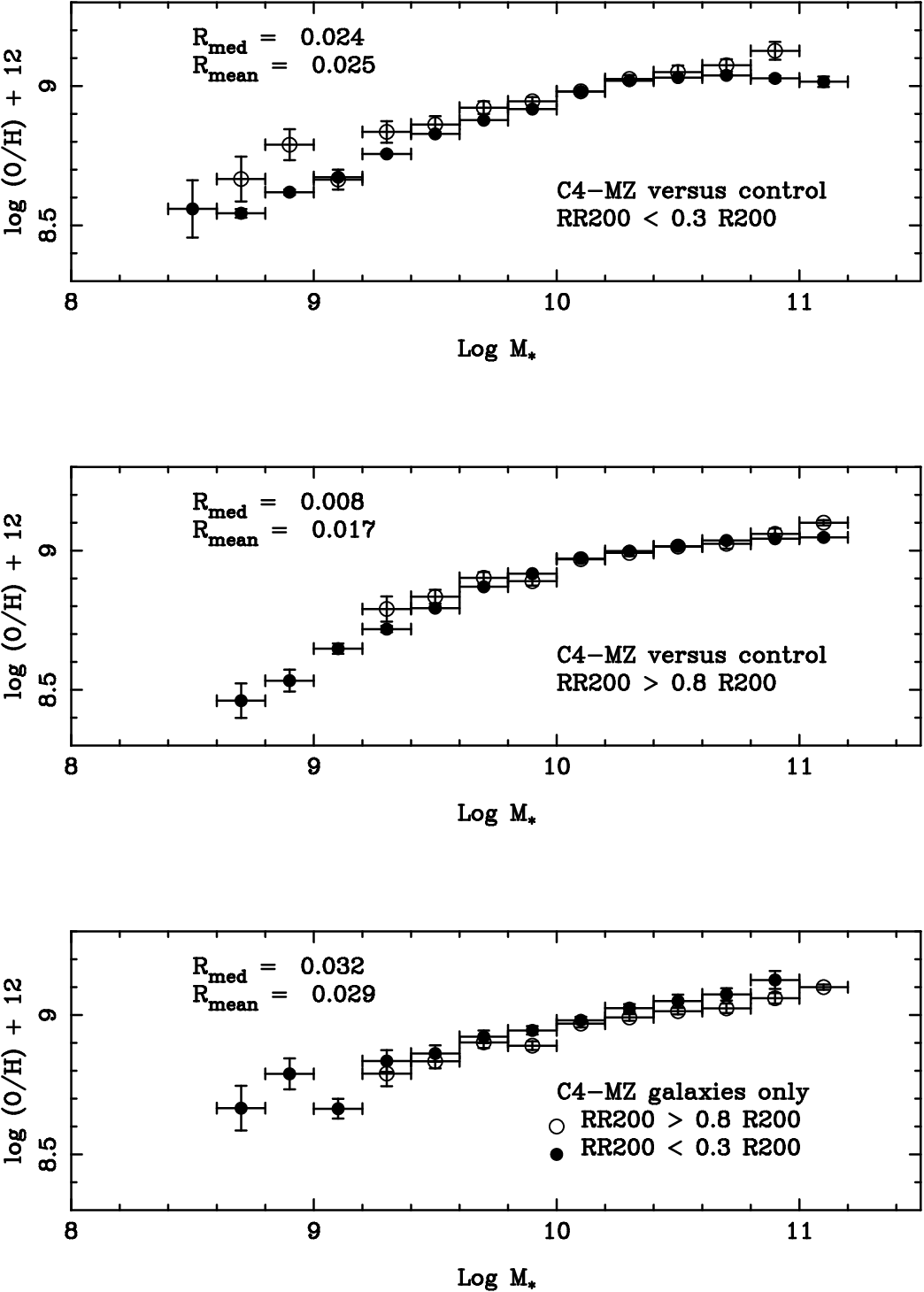}}}}
\caption{ The stellar mass-metallicity relation for C4-MZ cluster
galaxies (open circles) and their control galaxies (solid circles)
for two cuts in clustercentric distance (RR200) measured in units
of R200 (upper and middle panels).  The lowest panel shows the
mass-metallicity relation for only the C4-MZ galaxies for the same
cuts in RR200 (i.e. the open circles from the middle and upper panels).
\label{MZ_RR200}}
\end{figure}

Next we investigate whether cluster properties impact the
offset of the mass-metallicity relation.  The
C4-MZ galaxies are divided based on the $\sigma_v$, R200 and mass of its
host cluster.  In Figure \ref{plot_MZ_cluster_prop} these MZRs are compared,
in each case selecting the
control galaxies that were matched to the cluster galaxies
in each cut.  Figure  \ref{plot_MZ_cluster_prop} shows that
the global properties of the cluster do not strongly impact the stellar
mass-metallicity relation of the C4-MZ galaxies relative to
the control.  Visually, there may be a marginally enhanced metallicity
offset for clusters with smaller sizes and masses, but this effect
is not statistically significant.  In Figure \ref{MZ_RR200}
we show the MZR for C4-MZ galaxies
at small (RR200 $<$ 0.3 R200) and large (RR200 $>$ 0.8 R200)
clustercentric distances.  A small excess in metallicity in the cluster
galaxies is present in both subsets, although it is
slightly more pronounced at small values of RR200. There is
no correlation between RR200 and either r$_h$ or SSFR that
could be causing this difference.  We also show in the lowest
panel of Figure \ref{MZ_RR200} the MZR of
the C4-MZ clusters split by RR200.  There is an offset to higher median
nebular
metallicities by $\sim$ 0.03 dex in galaxies within 0.3 R200 relative
to galaxies at clustercentric distances above 0.8 R200.  However, these
two sub-samples in RR200 are not necessarily matched in mass, redshift
and $g-r$, as is the case for the control sample.  Nonetheless,
the results of Figure \ref{MZ_RR200} indicate that
the offset in the MZR may be regulated by local factors, such as position 
within the
cluster and not the simple fact of cluster membership; this possibility
is investigated further in the next section.

\section{Local environment}\label{env_sec}

\subsection{Cluster galaxies with a close companion}\label{pairs_sec}

\begin{figure}
\centerline{\rotatebox{0}{\resizebox{8cm}{!}
{\includegraphics{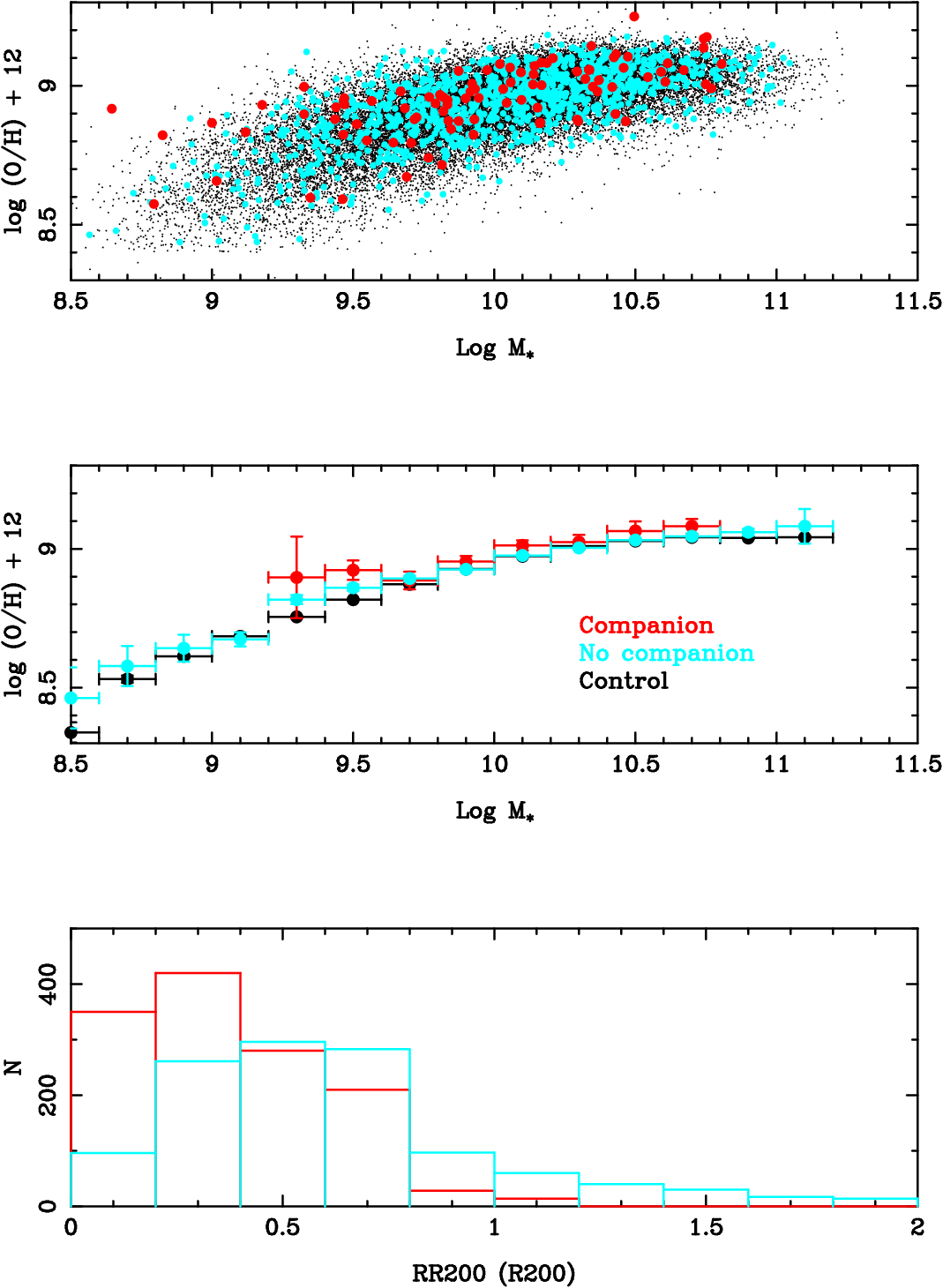}}}}
\caption{ Top panel: Mass-metallicity relation for C4-MZ galaxies with
(red) and without (cyan) a close companion within a
projected distance of 80 \hkpc\ and $\Delta v <$ 500 \kms.  Black
points indicate the control galaxies.  Middle panel: Same as
top panel, but with the MZR shown in bins of stellar mass.
Bottom panel:  Distribution of clustercentric radii for C4-MZ galaxies with
and without a close companion.  Colour coding as in top panel.
The histogram for galaxies with a close companion has been scaled up by
a factor of 13 for presentation purposes.
\label{pairs_MZ}}
\end{figure}

As a first test of the relative importance of local
environment in the normalization of the MZR, the C4-MZ
galaxies are cross-correlated with the sample of close pairs
presented in Ellison et al. (2008a).  As discussed by Barton et al. (2007),
a significant fraction of galaxies with a close companion occur in
clusters.  In brief, the pairs sample of Ellison et al.
contains 2887 galaxies with a companion within a 10:1 stellar mass ratio
range, projected separation $r_p <$ 80 \hkpc\ and $\Delta v < 500$ \kms.
95 galaxies (out of 1318) in the C4-MZ sample were matched to the 
close pairs sample, indicating that these 95 all have a close companion.
In Figure \ref{pairs_MZ} we compare the MZR of these 95 C4-MZ galaxies
with a close companion with the 1223 C4-MZ galaxies with no near neighbour 
and the 27678 control galaxies.  Both the binned and unbinned MZRs
are shown, although the close companion sample is only binned at
log M$_{\star} \ge 9.2$ M$_{\odot}$ due to small number statistics
at lower masses.  Figure \ref{pairs_MZ} shows that those C4-MZ galaxies
with close companions have a higher metallicity by 
$\sim$ 0.05 dex than
C4-MZ galaxies with no neighbour within $r_p$ = 80 \hkpc\ and $\Delta v$ =
500 \kms.  In fact, the C4-MZ galaxies without companions have metallicities
consistent with the control sample.  This is consistent with the result of 
Park et al. (2008) that companions within a few hundred kpc have a 
more important effect on morphology, star formation and luminosity than the
larger scale environment.The lower panel of Figure   \ref{pairs_MZ}
shows the distribution of clustercentric radii of the companion/no
companion samples.    Unsurprisingly, cluster galaxies with a near
neighbour reside preferentially at small values of RR200.

In general, the pairs sample is affected by fibre collisions; Figure
1 of Ellison et al. (2008a) shows these collisions affect completeness
as a function of redshift and galaxy separation.  Patton \& Atfield
(2008) find that whilst the overall spectroscopic completeness of the
sample is 88\%, for separations $<$ 55 arcsecs, this percentage
drops to $\sim$ 26\% (on average) at 
smaller separations.  It is therefore likely that there are galaxies
with close companions that are currently included among the 1223 non-pair
galaxies.  Indeed, Barton et al. (2007) used N-body simulations to show that
65\% of galaxy pairs are located in halos with a total of at least 4 galaxies.
They conclude that the majority of pair galaxies are located in the
group environment.  
The offset in the MZR between C4-MZ galaxies with and without close
companions may therefore be even stronger than shown in Figure \ref{pairs_MZ}.

\subsection{Local density estimators}

The effect of near neighbours can be further investigated by examining
the dependence of the MZR on local galaxy density.  Two
different density estimates are considered, and for each one the
MZR in the most extreme (rich and poor) environments is compared
to the matched control galaxies.

\subsubsection{The three-dimensional n density parameter}
 
Cowan \& Ivezic (2008) use a 3-dimensional density estimator
\footnote{Cowan \& Ivezic (2008)
actually use the equation n = C $\frac{1}{\sum_{i=1}^{10} d_i^3}$ where the
empirically determined value of C=11.48 allows the density parameter
$n$ to be translated into a physically meaningful number of galaxies
per Mpc${^3}$.} which sums the distances 
to the ten nearest galaxies to a given point in space (see Ivezic et al.
2005 for the full details of this formulism):

\begin{equation}\label{n10_eqn}
n_{10} = \frac{1}{\sum_{i=1}^{10} d_i^3}.
\end{equation}

When $n_{10}$ is calculated at the position of a galaxy, the latter counts 
as its own closest neighbour, so that $d_1 = 0$.
Therefore for our purposes we essentially only consider
the nine nearest neighbours:

\begin{equation}\label{n9_eqn}
n = \frac{1}{\sum_{i=1}^{9} d_i^3}.
\end{equation}

This estimator of density is attractive since it uses full
three-dimensional information to determine the volume density of
galaxies from a certain point.  However, it has two main drawbacks.
First, there is no correction for fibre collisions, which may lead to
an under-estimate of the local density in regions rich in projected
galaxies.  Second, the three-dimensional distances between galaxies do
not (can not) take the peculiar cluster motions into account.  The
significant velocity dispersions in clusters (see Figure
\ref{galaxy_metals_neb} for the $\sigma_v$ distribution in our sample)
will lead to significant errors in the radial distances
between cluster members (a manifestation of the so-called
fingers-of-god effect).  However, Cowan \& Ivezic (2008) show that
$n$ is fairly robust to this effect, being under-estimated by
up to 10\% at the typical redshift of our sample.  A third issue which
may be of concern is Malmquist bias: for a survey of given apparent
magnitude limit, lower redshift galaxies will appear to have more
faint neighbours than higher redshift galaxies, due to incompleteness.
This can be circumvented by imposing an absolute magnitude cut that
yields a volume limited sample.  For example, the density parameter
used by Mouhcine et al. (2007) imposes an absolute magnitude cut of
$M_r < -20$ which yields a volume-limited sample at $z < 0.085$ (see
also Baldry et al. 2006). Cowan \& Ivezic (2008) also apply cuts in
luminosity and redshift, but the raw values of $n$ are not volume
limited for our cluster sample.  However, as long as the C4-MZ sample
galaxies are compared only with their matched control galaxies,
Malmquist bias should be unimportant, since the two samples are
matched in redshift, colour and stellar mass.  That is, the equivalent
bias in luminosity and colour selection should apply equally to the 21
control galaxies matched to each C4-MZ galaxy.

\subsubsection{The two-dimensional $\Sigma$ density parameter}

Mouhcine et al. (2007) use a density estimator based on an average of
the \textit{projected} distances to the 4th and 5th nearest neighbour 
within 1000 \kms\ (see also Baldry et al. 2006):

\begin{equation}\label{sigma_eqn}
\log \Sigma = \frac{1}{2} \log \left(\frac{4}{\pi d_4^2}\right) + \frac{1}{2} \log \left(\frac{5}{\pi d_5^2}\right).
\end{equation}

For their paper, log $\Sigma$ was calculated only using neighbour
galaxies brighter than $M_r = -20$, in order to obtain a
volume-limited sample for $z<0.085$.  The C4 cluster galaxy sample
contains BCGs out to a redshift of $z=0.1$ and cluster members out to
$z=0.11$.  The $\Sigma$ densities are therefore re-calculated from Eqn
\ref{sigma_eqn} for $M_r < -20.6$ to yield a volume limited sample to
this higher redshift cut-off.  To account for missing spectroscopic
redshifts, log $\Sigma$ is also calculated using photometric
redshifts.  The final value of log $\Sigma$ is the average of the
spectroscopic redshift value and the value calculated with the
inclusion of photometric redshifts (see Baldry et al 2006 for further
discussion).  Although not directly affected by the fingers-of-god
effect, the calculation of $\Sigma$ still requires neighbour galaxies
to be within 1000 \kms.  The most massive clusters (largest
$\sigma_v$) in our sample may therefore have their $\Sigma$ values
under-estimated.  We repeated all of the metallicity analysis
presented in the following section for $\Sigma$ re-calculated for
galaxies within 2000 \kms.  As expected, the corresponding densities
were systematically higher, but they do not affect the conclusions of
the MZR analysis.  The 1000 \kms\ limit is therefore adopted.

\begin{figure*}
\centerline{\rotatebox{270}{\resizebox{12cm}{!}
{\includegraphics{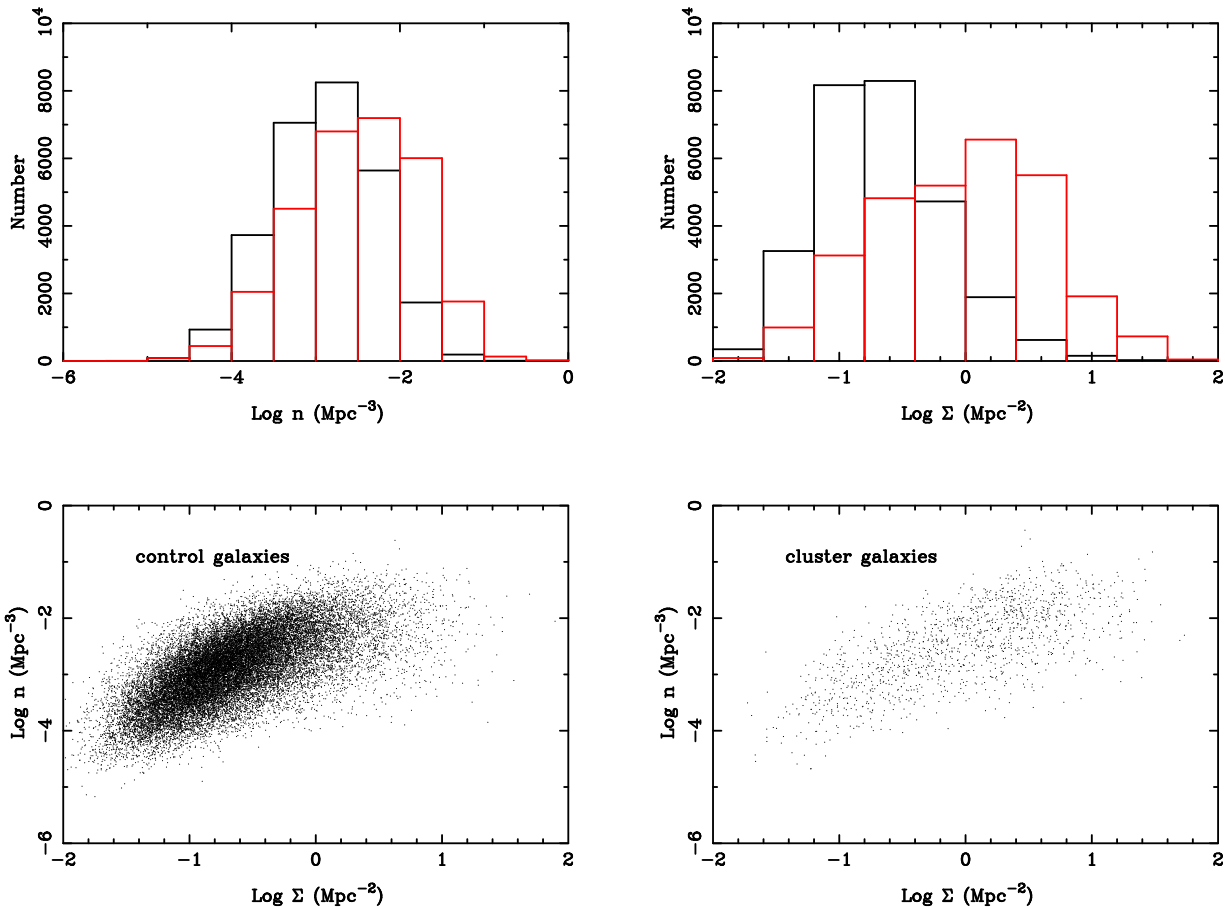}}}}
\caption{Top panels: The distribution of galaxy environments for the C4-MZ
sample (red) and the matched control sample (black) for two different
density estimators (see Eqns \ref{n9_eqn} and \ref{sigma_eqn}).
Bottom panels: A comparison of the galaxy densities for the $\Sigma$ and
$n$ estimators for the control sample (left) and C4-MZ sample (right).
\label{ivan_vs_nick}}
\end{figure*}

\begin{figure}
\centerline{\rotatebox{0}{\resizebox{8cm}{!}
{\includegraphics{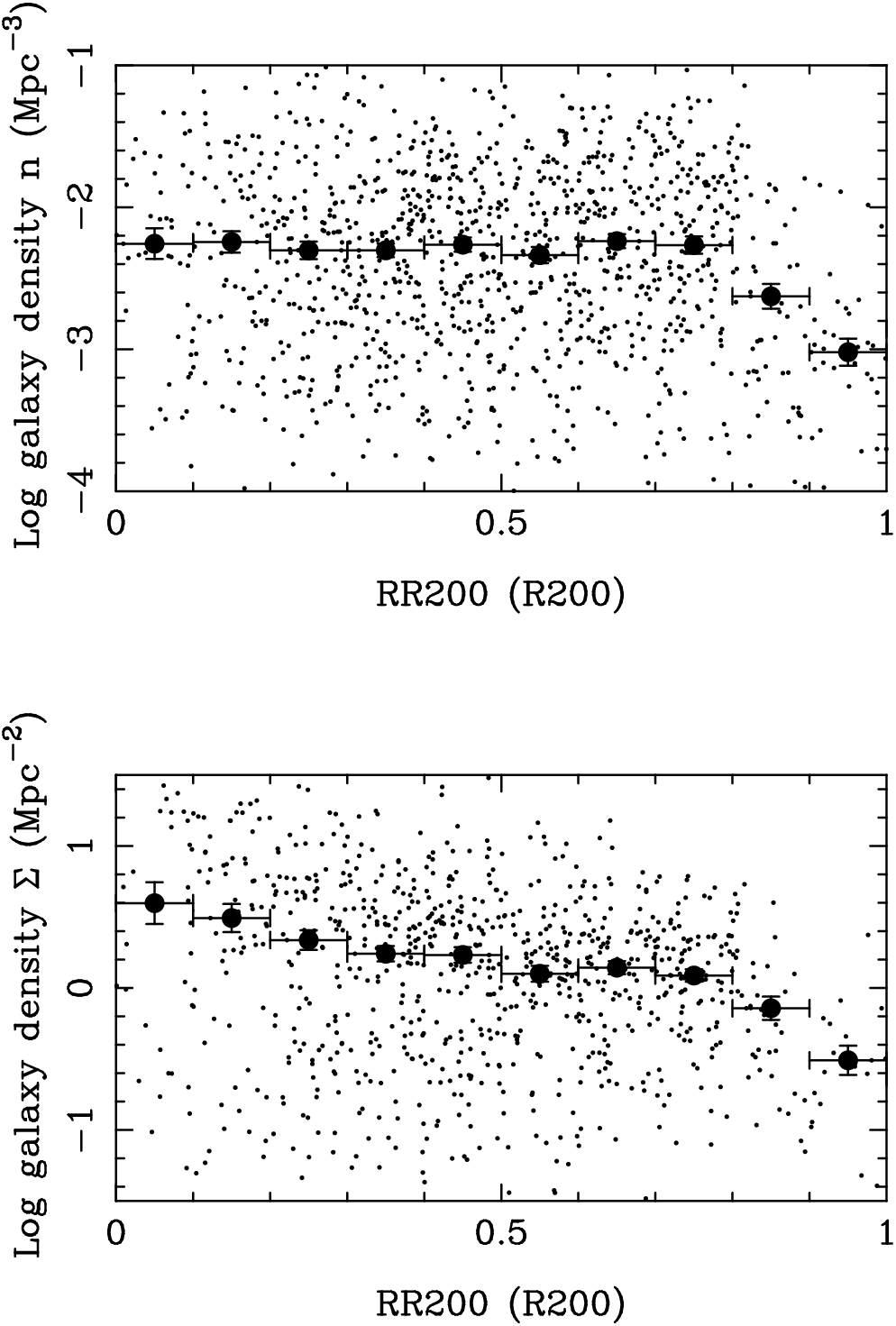}}}}
\caption{Galaxy environment parameters for the C4-MZ sample as
a function of clustercentric distance (RR200).  Individual
galaxy values as well as binned values are shown.  The upper panel
shows the density as defined by $n$ (Eqn \ref{n9_eqn}) and
the lower panel as defined by $\Sigma$ (Eqn \ref{sigma_eqn}).
\label{rr200_dens}}
\end{figure}

In Figure \ref{ivan_vs_nick}, the distribution of $n$ and $\Sigma$ is
shown for both the C4-MZ sample and their matched control galaxies.
For reference, we also compare the two density estimators for the cluster
and control galaxies.  In Figure \ref{rr200_dens} the two densities
are plotted as a function of clustercentric distance (RR200) for
the C4-MZ sample.  Although $\Sigma$ shows a clear dependence on RR200,
the relationship between $n$ and RR200 is very flat.  This may be
indicative of a combination of the fingers-of-god and fibre incompleteness
discussed in the previous section, which lead to an under-estimate
of $n$ in the cluster environment.

\subsection{The impact of local environment on the mass-metallicity relation}

\begin{figure*}
\centerline{\rotatebox{270}{\resizebox{12cm}{!}
{\includegraphics{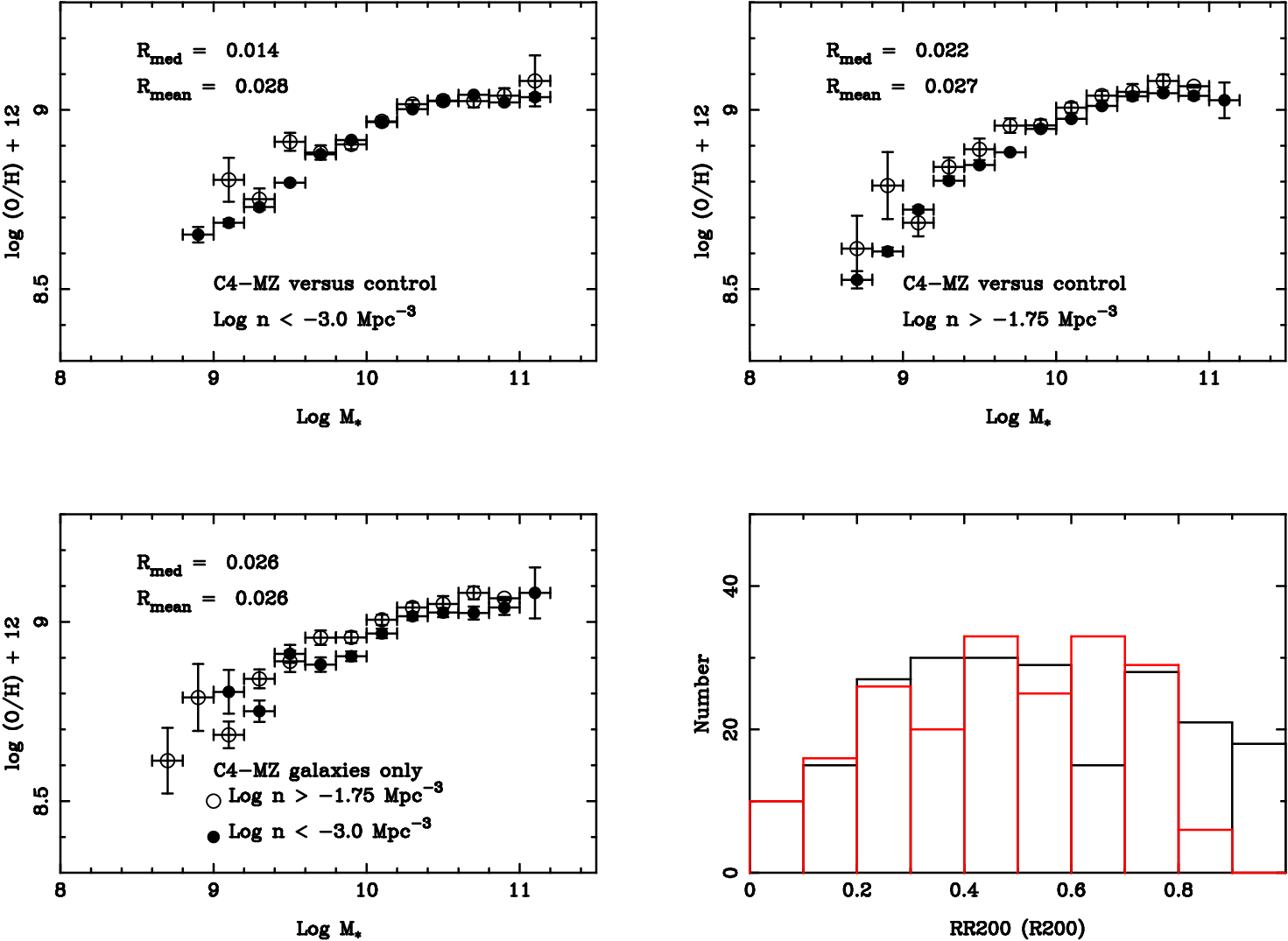}}}}
\caption{The mass-metallicity relation for different cuts in local density,
as measured by $n$.
In the top panels, solid points are the control galaxies
and open points are the C4-MZ galaxies.  In the bottom left panel, only
C4-MZ galaxies are shown for the same two density cuts (i.e. the open
points from the top-right and top-middle panels).   The lower right panel
shows the clustercentric distances for the C4-MZ galaxies for the
two density cuts: log n $< -3$ in black and  log n $> -1.75$ in red.
\label{MZR_envdens_n}}
\end{figure*}

\begin{figure*}
\centerline{\rotatebox{270}{\resizebox{12cm}{!}
{\includegraphics{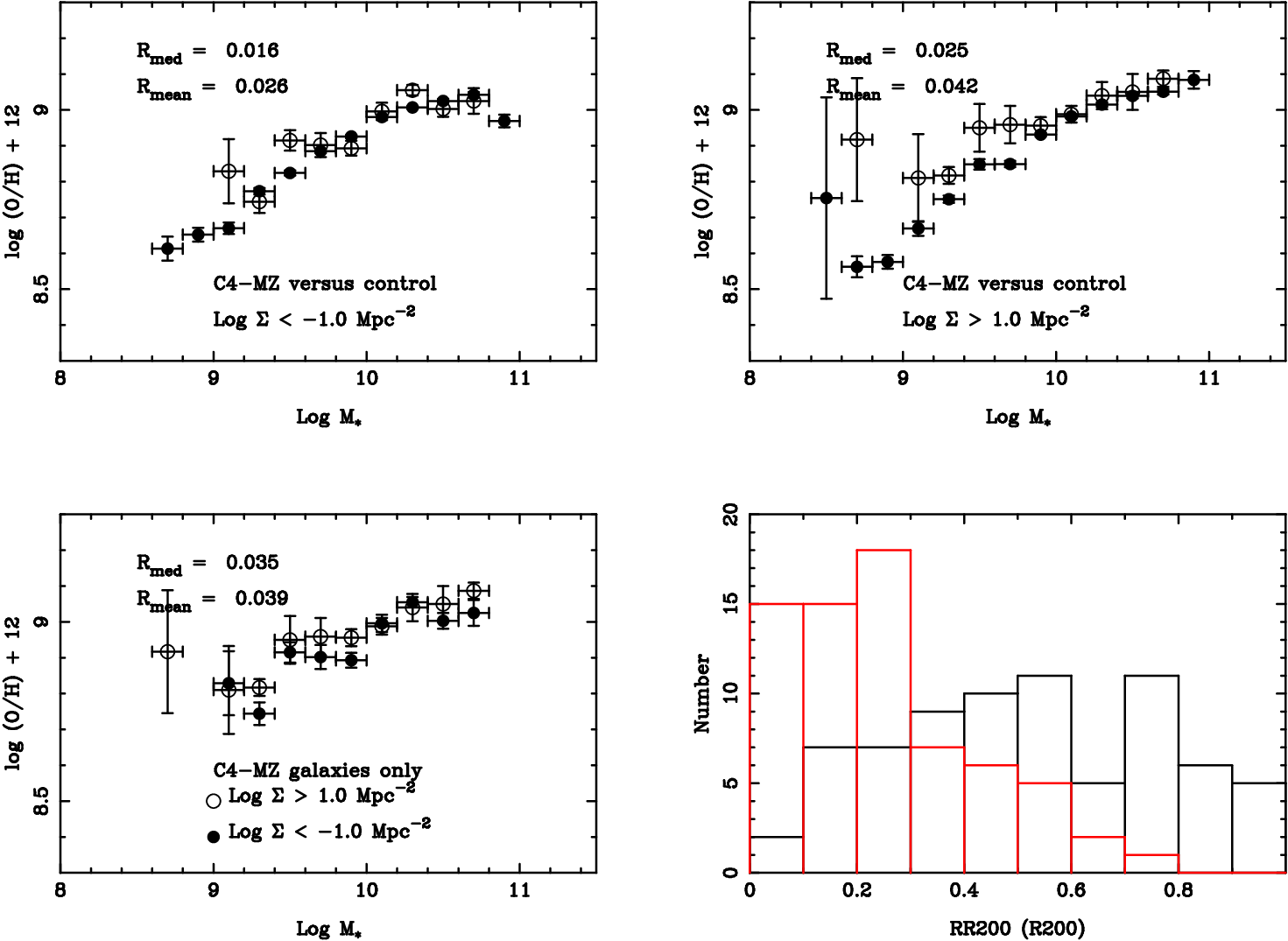}}}}
\caption{The mass-metallicity relation for different cuts in local density,
as measured by $\Sigma$.
In the top panels, solid points are the control galaxies
and open points are the C4-MZ galaxies.  In the bottom left panel, only
C4-MZ galaxies are shown for the same two density cuts (i.e. the open
points from the top-right and top-middle panels).   The lower right panel
shows the clustercentric distances for the C4-MZ galaxies for the
two density cuts: log $\Sigma < -1.0$ in black and  log $\Sigma > 1.0$ in red.
\label{MZR_envdens_sigma}}
\end{figure*}

In Section \ref{C4_MZR_sec} we showed that galaxies in the C4-MZ
sample have slightly higher metallicities at a given stellar mass
compared to the control sample.  We also showed that the offset in
the MZR does not appear to depend on global cluster properties (R200, 
$\sigma_v$, cluster mass), but that the offset is more pronounced
at small clustercentric distances.  There is also a strong offset
to higher metallicities when a C4-MZ galaxy has a close companion. 
These results indicate that
the offset in the MZR may be driven by local effects within the
cluster.

As shown in Figure \ref{ivan_vs_nick} cluster galaxies inhabit a wide
range of local environment densities, as measured by both $n$ and $\Sigma$.
We divide the C4-MZ galaxies into those that reside in particularly
rich local environments (log $n > -1.75$ Mpc $^{-3}$, log $\Sigma > 1.0$ 
Mpc $^{-2}$) and those in relatively poor environments (log $n < -3.0$ 
Mpc $^{-3}$, log $\Sigma < -1.0$ Mpc $^{-2}$).    In Figures 
\ref{MZR_envdens_n} and \ref{MZR_envdens_sigma}
the MZR of the C4-MZ galaxies in these environmental extremes
are compared to their matched controls for the two different
density estimators.  These Figures show that the offset towards
high metallicities for a given mass is more pronounced in the
high density environments, as measured by either $n$ or $\Sigma$.
Indeed, for $n$ (Figure \ref{MZR_envdens_n}), the MZR of cluster
galaxies in locally poor environments is actually consistent
with the control sample.  The dependence on environment
is emphasized in the lower left panels of Figures 
\ref{MZR_envdens_n} and \ref{MZR_envdens_sigma} where the metallicities
of C4-MZ galaxies in locally rich environments (open points)
are systematically higher than cluster galaxies in low density
environments (filled points).

Is the dependence of the MZR on local environment a simple reflection
of a dependence on clustercentric distance (e.g. Figure
\ref{MZ_RR200})?  For example, in Section \ref{pairs_sec} it was shown
that although C4-MZ galaxies with a close companion are more
metal-rich than both the cluster galaxies with no near neighbour, and
the control, they also preferentially reside at lower clustercentric
distances.  A similar effect is seen in Figure \ref{MZR_envdens_sigma}
where the distribution of RR200 is clearly skewed to low values for
C4-MZ galaxies in high density environments.  This is not surprising
given the anti-correlation between $\Sigma$ and RR200 previously shown
in Figure \ref{rr200_dens}.  Conversely, for $n$, the distribution of
RR200 shown in the lower right panel of Figure \ref{MZR_envdens_n} is
very similar for the C4-MZ galaxies in both rich and poor
environments.  As previously discussed, spectroscopic incompleteness
due to fibre collisions leads to an under-estimate of $n$ in the
densest environments (such as cluster cores), resulting in a flat
relationship between RR200 and $n$ (see Figure \ref{rr200_dens}).
Therefore, the high and low $n$ samples are probing the same wide
range of RR200, so the offset between the two mass-metallicity
sequences in Figure \ref{MZR_envdens_n} cannot be caused by a
dependence on clustercentric distance.

In order to further differentiate between a dependence on RR200 and local
density, we calculate the polynomial fit to the MZR of the control
galaxies.  We derive a good fit with a cubic polynomial of the form

\begin{equation}\label{MZR_fit_eqn}
 \begin{array}{ll} \log (O/H) + 12 = & 42.243 - 11.6452 \times 
\log M_{\star} \\
& + 1.30731 \times (\log M_{\star})^2 \\
& - 0.047577 \times (\log M_{\star})^3.
\end{array}
\end{equation}

\begin{figure}
\centerline{\rotatebox{0}{\resizebox{8cm}{!}
{\includegraphics{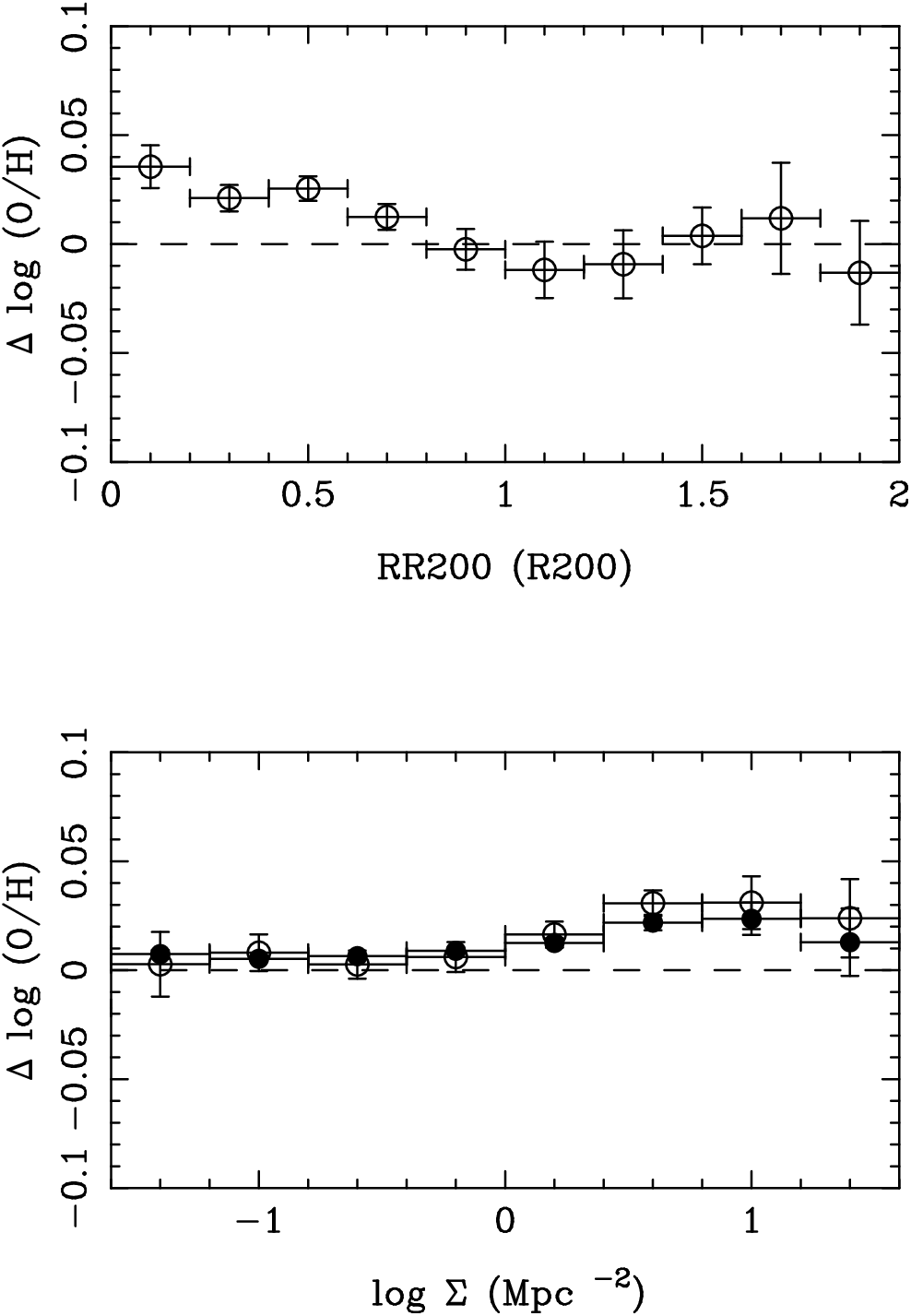}}}}
\caption{$\Delta$ log (O/H) is the difference between the measured
galaxy metallicity and the polynomial fit to the control MZR given in
Eqn \ref{MZR_fit_eqn}.  The metallicity offset is plotted as a
function of local environment density.  Open symbols represent C4-MZ
cluster galaxies and solid symbols are the control sample.  The
positive difference for control galaxies at high local densities
indicates that cluster membership is not the dominant factor in
driving the cluster galaxies to enhanced metallicities.
\label{dOH}}
\end{figure}

The $\Delta$ log (O/H) of each galaxy in both the cluster
and control samples is then calculated as the difference between the
measured oxygen abundance and the value inferred from the stellar mass
inserted into Eqn. \ref{MZR_fit_eqn}.  In Figure \ref{dOH} we plot
the values of $\Delta$ log (O/H), i.e. the deviation from the best
fit control MZR, as a function of RR200 (C4-MZ sample only) and $\Sigma$
(C4-MZ and control samples).  The top panel of  Figure \ref{dOH}
re-iterates the previous result that the offset of cluster galaxies
from the MZR depends on clustercentric distance.  This figure shows
more clearly that enhanced metallicities are present at RR200 $\lesssim$ 0.7 
R200.  At the smallest clustercentric distances, the median metallicity
in the C4-MZ sample is
higher than the fit to the control MZR by $\sim$ 0.035 dex.
In the bottom panel of  Figure \ref{dOH} we plot the  $\Delta$ log (O/H)
for both cluster and control galaxies.  As shown in Figure 
\ref{ivan_vs_nick}, both the cluster and control galaxies span
a similar range in local density (as defined by either $n$ or $\Sigma$).
Both the C4-MZ cluster galaxies and the control (non-cluster) sample
show enhanced median metallicities by up to 0.04 dex for log $\Sigma> $ 0.  
This indicates that local density, even outside the cluster environment,
can drive the MZR to higher metallicities.  Interestingly, the cluster
galaxies are consistently (although not significantly) more metal-rich
at local densities log $\Sigma> $ 0.  This may indicate that cluster
membership has a minor, secondary effect on a galaxy's metallicity.

\begin{figure}
\centerline{\rotatebox{270}{\resizebox{6cm}{!}
{\includegraphics{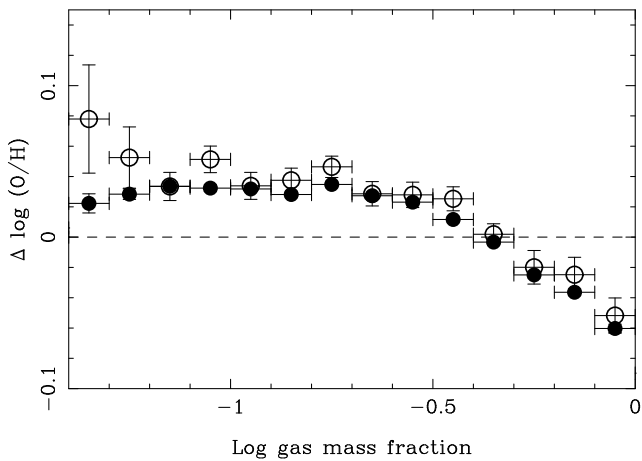}}}}
\caption{$\Delta$ log (O/H) is the difference between the measured
galaxy metallicity and the polynomial fit to the control MZR given in
Eqn \ref{MZR_fit_eqn}.  The metallicity offset is plotted as a
function of inferred HI-to-stellar mass ratio.  Open symbols represent C4-MZ
cluster galaxies and solid symbols are the control sample.  Gas-poor
galaxies in both the cluster and control samples show positive
metallicity offsets towards lower gas fractions.
\label{fgas_fig}}
\end{figure}

\section{Summary and discussion}

Using a sample of 1318 cluster galaxies with reliable nebular
metallicities, we have investigated how the stellar
mass - gas phase metallicity relation
responds to environmental effects.  The main conclusions of this
study are:

\begin{enumerate}

\item On average, galaxies in clusters are more metal rich than a
control sample matched in mass, $g-r$ colour, redshift and fibre
covering fraction by up to 0.04 dex.  A similar offset is seen in the
luminosity-metallicity relation.

\item The offset to higher metallicities for cluster galaxies is
independent of the global cluster properties (R200, $\sigma_v$ and
cluster mass).  However, galaxies at small clustercentric radii (RR200
$<$ 0.3 R200) have median metallicities that are higher than galaxies
at large clustercentric radii (RR200 $>$ 0.8 R200) by 0.03 dex.

\item  Cluster galaxies with a close companion have a higher
median metallicity  than cluster galaxies without a close companion
by $\sim$ 0.05 dex.

\item Two parameterizations were used to quantify the local density
of galaxies in clusters: a three-dimensional estimator, $n$, and a 
two-dimensional estimator, $\Sigma$.  Both density estimators
indicate that the metallicity offset between clusters and the matched
control sample is larger for richer environments.  The difference
in median metallicity between cluster galaxies in the richest and poorest
environments at a given stellar mass is 0.02 -- 0.07 dex.

\item  Both cluster and control galaxies show a metallicity dependence
on local environment.

\end{enumerate}

Taken together, these results indicate that although cluster galaxies
have higher average metallicities at a given stellar mass, neither simple
cluster membership, nor cluster properties seem to drive this effect.
Instead, it is apparently local scale processes, such as the presence
of a close companion or several near neighbours that leads to an
enhanced metallicity.  These results are similar in nature to previous
findings that galaxy properties are independent of cluster mass, but
do depend on clustercentric distance or local overdensity (e.g. Balogh
et al. 2004; Martinez, Coenda \& Muriel 2008).  The median metallicity
offset between cluster and control galaxies is only 0.04 dex, which
may explain why previous studies of cluster dwarfs found no clear
offset in the MZR with respect to field dwarfs (Vilchez 1995; Lee et
al. 2003).  The metallicity offset is so subtle that unbinned data
(even in a large sample, e.g. Figure \ref{plot_MZ_unbin_rh}) does not
clearly demonstrate the metallicity enhancement.

Our conclusions agree with those drawn by Skillman et al. (1996) and
Dors \& Copetti (2006) who found metal enhancements for (some)
individual galaxies in the Virgo cluster.  The statistics of the
current study are far superior to that of the Virgo cluster studies
which only included 9 cluster galaxies.  The binned mass-metallicity
relations presented here are therefore much less affected by scatter.
However, our results are most powerful in probing the \textit{average}
metallicity offset at a given stellar mass, which is distinct from
assessing the way a given galaxy responds to its environment.  This
subtlety is highlighted by the finding of Skillman et al. (1996) and
Dors \& Copetti (2006) that it is only the gas-deficient Virgo cluster
galaxies that show metallicity enhancements.  In these cases, the
metallicity enhancement is much larger (typically $\sim$ 0.3 dex) than
our average metallicity offset, even at the highest local
overdensities.  Zhang et al. (2009) have recently shown that for all
star-forming galaxies in the SDSS DR4, there is a tendency for
gas-deficient galaxies to exhibit higher metallicities at a given
stellar mass.  Ellison et al. (2008b) also showed that metallicities
are high for a given stellar mass for the galaxies with the lowest
SSFRs.  The low gas fractions are consistent with the picture that
previously efficient star formation has depleted the gas and converted
it to stars, leaving a relatively low rate of star formation at the
present time.

Gas fractions are not available for the majority of our galaxy sample,
although the ALFALFA survey will eventually map approximately 4000
square degrees of the SDSS survey area, covering redshifts out to $z
\sim$ 0.06.  As an alternative measure of gas fraction, we use the
calibration of Zhang et al. (2009) which is based on SDSS $g-r$ colour
and galaxy surface brightness in the $i$-band.  In Figure
\ref{fgas_fig} the metallicity offset from the control MZR is shown as
a function of the ratio of inferred HI-to-stellar mass for both the
C4-MZ and control samples.  A notable caveat here is that the
application of this calibration to cluster galaxies has not yet been
tested, so the assumption here is that galaxy colour and surface
brightness respond to gas fraction equally in all environments.
Enhanced metallicities are seen in both the cluster and control
samples when the inferred gas fraction is low, as expected from the
results of Zhang et al. (2009).  However, the cluster galaxies show an
additional systematic metallicity enhancement over the field galaxies.
This may be connected with the tentative result of Figure \ref{dOH}
that cluster galaxies are more metal-rich than the field at a given
over-density.  For example, if the richest environments suffer gas
exhaustion through both star formation and stripping (ram
pressure/strangulation), e.g. Chung et al. (2007).

It has been debated (e.g. Balogh et al. 1997 and references therein)
whether or not galaxies undergo a burst of star formation as they
enter the cluster environment which quenches further activity, or
whether star formation is simply shut-down, e.g. by gas stripping.
The result that galaxies in clusters have a slightly higher
metallicity for a given stellar mass than a matched control sample may
seem to support the idea of cluster-triggered star formation.
However, this is not the only interpretation.  If gas is ram pressure
stripped from cluster galaxies promptly after entering the cluster,
then the deposition of metals into the ISM from the last generation of
stars will lead to a relatively high nebular metallicity.  The
selection of our galaxy sample to have strong emission lines means
that they do still contain gas, confirmed by our estimates of gas
fraction (see also Tremonti et al. 2004 for alternative derivations of
the gas fraction).  Simulations indicate that ram pressure stripping
can remove up to 80\% of a galaxy's gas within 10$^7$ years (Abadi et
al.  1999), although most cluster galaxies undergo more modest gas
removal (McCarthy et al. 2008).  It is therefore interesting that a
similar metallicity enhancement is seen in galaxies in high local
densities (as measured by $\Sigma$), but outside of the cluster
environment.  Ram pressure stripping can remain effective even in
groups of galaxies (e.g. McCarthy et al.  2008; Sengupta et al. 2007)
and tidal forces can disturb or strip gas in more isolated
environments.  Moreover, as noted in Section \ref{met_sec} in both the
field and cluster environments high SSFR does not lead to enhanced
metallicities at a given mass.  The tendency of our control sample
galaxies to have enhanced metallicities, but at a slightly lower level
than the cluster galaxies at the same local density and gas fraction,
may reflect that both gas depletion through star formation and some
stripping process contribute to gas exhaustion in clusters.

Peeples et al. (2008) studied a sample of low mass, high metallicity
outliers in the MZR which reside in a similar parameter space to some
of the C4-MZ galaxies with close companions.  However, the metal-rich
dwarfs in the Peeples et al. sample are relatively isolated, with no
close companions and are not in clusters.  Peeples et al. (2008)
conclude that these objects are likely to be transitional dwarfs with
low gas fractions that are nearing the end of their star forming
lifetimes.  Based on the finding that gas-rich and gas-poor dwarfs
follow the same MZR, Lee, Bell \& Somerville (2008) have argued that
most dwarf spheroidals are formed by the removal of gas, a process
linked to local environment.  However, since the metal-rich outliers
of Peeples et al.  (2008) are morphologically consistent with dwarf
spheroidals, yet have no near neighbours and are not located in
clusters, these galaxies may have avoided gas stripping and hence been
able to evolve peacefully and achieve their relatively high
metallicities.  It therefore appears that there may be more than one
route to getting (metal) rich.

\section*{Acknowledgments} 

We are grateful to Anja von der Linden, Lisa Kewley and Christy
Tremonti for providing their SDSS data tables and to Evan Skillman and
Wei Zhang for useful comments.  This research
also made extensive use of the MPA/JHU public SDSS data tables
(http://www.mpa-garching.mpg.de/SDSS/) which provide a wonderful
community resource.
SLE and DRP acknowledge the receipt of NSERC Discovery grants which
funded this research.

\end{document}